\documentclass[aps,pra,onecolumn,superscriptaddress,longbibliography,nofootinbib,showkeys,showpacs]{revtex4-1}
\usepackage{blkarray}
\usepackage{amsfonts}
\usepackage{amsmath}
\usepackage{graphicx}
\usepackage{appendix}
\usepackage{bm}
\usepackage{float}
\usepackage{braket}
\usepackage{url}
\usepackage{xcolor}
\usepackage{comment}
\usepackage[top=2cm,left=2cm,right=2cm,bottom=2cm]{geometry}
\usepackage{hyperref}
\usepackage[utf8]{inputenc}
\usepackage{mathptmx}
\usepackage{CJK}
\usepackage[capitalise]{cleveref}
\usepackage{cases}
\usepackage[ruled,vlined,linesnumbered,boxed]{algorithm2e}%[ruled,vlined]{

\begin{document}
\colorlet{CV}{.}
\SetKwRepeat{Do}{do}{while}
\title{Variational quantum algorithms for Poisson equations based on the decomposition of sparse Hamiltonians}

\author{Hui-Min Li}
%\email{lihuimin@cnu.edu.cn}
\affiliation{School of Mathematical Sciences, Capital Normal University, 100048, Beijing, China}
\author{Zhi-Xi Wang}
\email{wangzhx@cnu.edu.cn}
\affiliation{School of Mathematical Sciences, Capital Normal University, 100048, Beijing, China}
\author{Shao-Ming Fei}
\email{feishm@cnu.edu.cn}
\affiliation{School of Mathematical Sciences, Capital Normal University, 100048, Beijing, China}
%\date{\today}

\begin{abstract}
Solving a Poisson equation is generally reduced to solving a linear system
with a coefficient matrix $A$ of entries $a_{ij}$, $i,j=1,2,...,n$,
from the discretized Poisson equation. Although the variational quantum algorithms
are promising algorithms to solve the discretized Poisson equation,
they generally require that $A$ be decomposed into a sum of $O[\text{poly}(\text{log}_2n)]$ simple operators in order to evaluate efficiently the loss function. A tensor product decomposition of $A$ with $2\text{log}_2n+1$ terms has been explored in previous works. In this paper, based on the decomposition of sparse Hamiltonians we greatly reduce the number of terms. We first write the loss function in terms of the operator $\sigma_x\otimes A$ with $\sigma_x$ denoting the standard Pauli operator. Then for the one-dimensional Poisson equations with different boundary conditions and for the $d$-dimensional Poisson equations with Dirichlet boundary conditions, we decompose $\sigma_x\otimes A$ into a sum of at most 7 and $(4d+1)$ Hermitian, one-sparse, and self-inverse operators, respectively. We design explicitly the quantum circuits to evaluate efficiently the loss function. The decomposition method and the design of quantum circuits can also be easily extended to linear systems with Hermitian and sparse coefficient matrices satisfying $a_{i,i+c}=a_{c}$ for $c=0,1,\cdots,n-1$ and $i=0,\cdots,n-1-c$.
\end{abstract}

\maketitle

\section{Introduction}
Variational quantum algorithms (VQAs) have emerged as the leading strategy to attain quantum advantage on noisy intermediate-scale Quantum devices \cite{cerezo_variational_2021,Preskill_2018,Peruzzo_2014}. VQAs are a class of hybrid quantum-classical algorithms which have been successfully applied to a plethora of applications such as solving nonlinear problems \cite{Lubasch_2020}, linear systems \cite{Xu_03898,Huang_07344,harrow_quantum_2009} and combinatorial optimization problems \cite{farhi_quantum_2014,wurtz_maxcut_2021,crooks_performance_2018}.

The Poisson equation has fundamental importance in numerous areas of science and engineering, such as quantum mechanical continuum solvation \cite{Cao_2013,Tomasi_2005}, computational fluid dynamics \cite{Batchelor_2000} and the theory of Markov chains \cite{meyn_2007,meyn_tweedie_glynn_2009}. Driven by its importance, there is immense interest in solving the Poisson equation by using VQAs recently \cite{liu_variational_2021}. The main idea is to first discrete the Poisson equation to a linear system with a coefficient matrix $A$ by using the finite-difference method \cite{FGEW2008}, and then approximate the solution to the linear system through VQAs.

However, the application of VQAs on the linear system always requires that the coefficient matrix $A$ be decomposed into a sum of $O[\text{poly}(\text{log}_2n)]$ simple operators to evaluate efficiently the loss function, where $n$ is the dimension of $A$. For general linear systems, to find a strategy  satisfying the above requirements is a nontrivial problem \cite{liu_variational_2021}. Due to the special structure of the discretized Poisson equations, the authors in \cite{liu_variational_2021} find an explicit tensor product decomposition of $A$ under a set of simple operators $\{I,\,\sigma_{+}=|0\rangle\langle1|,\,\sigma_{-}=|1\rangle\langle0|\}$, for which the number of terms is $2\log_2 n+1$. The observables are constructed to evaluate the expectation values of these simple operators.

In this paper, inspired by the idea introduced in \cite{kirby_variational_2021}, we present a different decomposition strategy to greatly reduce the number of the decomposition terms in order to apply VQAs.
We first write the loss function in terms of the operator
$\sigma_x\otimes A$ with $\sigma_x$ denoting the standard Pauli operator, as $\sigma_x\otimes A$ can be directly decomposed into a sum of Hermitian, one-sparse and self-inverse operators. Then, for the one-dimensional Poisson equations with different boundary conditions and for the $d$-dimensional Poisson equations with Dirichlet boundary conditions, we decompose $\sigma_x\otimes A$ into a sum of at most 7 and $(4d+1)$ Hermitian, one-sparse and self-inverse operators, respectively. Finally, the quantum circuits are explicitly constructed to evaluate efficiently the loss function.
Although the idea of introducing ancilla qubits for sparse matrix decomposition with the number of decomposition terms being independent of the dimension $n$ has been proposed in \cite{kirby_variational_2021, berry_exponential_2014}, the queries to decomposition terms are usually referred to as black box. Thus, our method has practical significance as we adapt the decomposition strategy for the discretized Poisson equation and construct exactly the quantum circuits to realize the queries.

Our algorithm is effective as the number of the decomposition terms is only polynomial of the dimension $d$ of the Poisson equation, and is independent of $n$. Therefore, our algorithm greatly reduces the number of measurements. This advantage is particularly outstanding in dealing with the linear systems with large dimension $n$. In addition, the number of the ancilla qubits required for realizing the quantum circuits corresponding to the decomposition terms is at most $\log_2n+2$ and is independent of $d$. The circuit complexity is also polynomial of $\log_2n$ and independent of $d$, which implies that the benefit of our algorithm is still obvious for the high-dimensional Poisson equations. Moreover, the decomposition method and the quantum circuit design are still suitable and efficient for linear systems with Hermitian and sparse coefficient matrices satisfying $a_{i,i+c}=a_{c}$ for $c=0,1,\cdots,n-1$ and $i=0,\cdots,n-1-c$, where $a_{i,i+c}$ denotes the element of the coefficient matrices. Additionally, inspired by our method, for the coefficient matrices which slightly violate the above conditions one may also find the corresponding decompositions and circuit designs.

The paper is structured as follows. In Sec.~\ref{sec:HAMILTONIANS}, we briefly review the definition of Poisson equations, the linear system generated by discretizing the Poisson equations, and the Hamiltonian whose ground state encodes the solution to the linear system. In Sec.~\ref{sec:VQA_SPARSE_HAMILTONIAN}, we demonstrate how to approximate the ground state of the Hamiltonian by using VQAs based on the decomposition of sparse Hamiltonians. In Sec.~\ref{sec:Experiments}, we numerically illustrate the performance of our algorithm for the one-dimensional Poisson equations with Dirichlet boundary conditions. We discuss our results and conclude in Sec.~\ref{conclusion}.

\section{HAMILTONIAN FOR DISCRETE POISSON EQUATIONS}
\label{sec:HAMILTONIANS}
%In this section, we review the linear systems generated by discretizing the Poisson equations
%%one-dimensional Poisson equation with different boundary conditions and the d-dimensional Poisson equation with the Dirichlet boundary conditions,
%and the Hamiltonians whose ground states encode the solutions to the resulting linear systems.

We first review the definition of Poisson equations. The one-dimensional Poisson equations with different boundary conditions can be written as \cite{liu_variational_2021,Serigne2014_309,Serigne2014_372,ELC1998}
\begin{equation}
\begin{aligned}\label{one_dimensioanl_Poisson_equation}
\begin{split}
&-\bigtriangleup \mu(x)=f(x),~x\in (0,1),
\end{split}
\end{aligned}
\end{equation}
with the unified boundary conditions $\alpha_{1}\mu'(0)-\alpha_{2}\mu(0)=0$ and
$\beta_{1}\mu'(1)+\beta_{2}\mu(1)=0$, where $\bigtriangleup$ is the Laplace operator; $\alpha_{1},\alpha_{2},\beta_{1},$ and $\beta_{2}$ are all positive constants; and $f: D\rightarrow R$ is a sufficiently smooth function. The boundary conditions are given by $\alpha_{1},\alpha_{2},\beta_{1},$ and $\beta_{2}$. It corresponds to the Dirichlet boundary condition when $\alpha_{1}=0,\alpha_{2}=1,\beta_{1}=0$, and $\beta_{2}=1$.
For the $d$-dimensional Poisson equations, we consider the case with Dirichlet boundary conditions , namely, $-\bigtriangleup \mu(x)=f(x),x\in (0,1)^{d}$ with the boundary condition $\mu(x)=0$, $x\in\{0,1\}^{d}$.

 Aiming to solve the Poisson equations numerically, the finite-difference method is used to discretize Poisson equations to generate linear systems \cite{liu_variational_2021,FGEW2008}. Specifically, solving the one-dimensional Poisson equation with the unified boundary conditions can be transformed to the problem of solving the following linear system:
\begin{equation}\label{one_dimensional_linear_system}
\widehat{A}\mathbf{x}=\mathbf{b},~ \mathbf{b}\in R^{n}
\end{equation}
with
\begin{equation}
\begin{aligned}
\widehat{A}=\left[
\begin{array}{cccccc}
2-c  & -1  \\
-1   & 2  & -1  \\
     & -1 & 2      & -1 \\
     &    & \ddots & \ddots & \ddots\\
     &    &        & -1     & 2  & -1 \\
     &    &        &        & -1 & 2-d
 \end{array}
 \right]\in R ^{n\times n},
 \end{aligned}
 \end{equation}
where $c=\frac{\alpha_{1}}{\alpha_{1}+\alpha_{2}h}$, $d=\frac{\beta_{1}}{\beta_{1}+\beta_{2}h}$, $h=1/(n+1)$, $n$ comes from evenly dividing $(0,1)$ into $n+1$ parts during the discretization, and $\mathbf{b}$ is the vector obtained by sampling $f(x)$ on the interior grid points.

Similarly, the $d$-dimensional Poisson equation with the Dirichlet boundary conditions can be discretized to be the following linear system,
\begin{equation}\label{d_dimensional_linear_system}
\begin{aligned}
A^{(d)}\mathbf{x}&=\mathbf{b},\quad \mathbf{b}\in R^{n^d},\quad A^{(d)} \in R^{n^d \times n^d}
\end{aligned}
\end{equation}
with
\begin{equation}
\begin{aligned}
A^{(d)}&=\underbrace{\tilde{A}\otimes I \otimes \cdots \otimes I}_{d}+\underbrace{I\otimes \tilde{A}\otimes I \otimes \cdots \otimes I}_{d}+\cdots+\underbrace{I \otimes \cdots \otimes I\otimes \tilde{A}}_{d},\\
\mbox{where}\
\tilde{A}&=
\left[
\begin{array}{cccccc}
2  & -1  \\
-1   & 2  & -1  \\
     & \ddots & \ddots      & \ddots \\
     &    &        & -1     & 2  & -1 \\
     &    &        &        & -1 & 2
 \end{array}
 \right]\in R ^{n\times n},~~I \in R^{n\times n}.
\end{aligned}
\end{equation}

Note that the error between the exact solution and the numerical solution obtained by solving the corresponding linear system is $O(1/n^2)$ \cite{Yoon2015}. In other words, the fidelity between the exact solution and the numerical solution increases with the increase of $n$. However, the computation of the linear system can be inefficient for large $n$.

In order to compute the linear systems \eqref{one_dimensional_linear_system} and \eqref{d_dimensional_linear_system} efficiently for large $n$, the problem of solving the linear system is transformed into finding the unique ground state of the Hamiltonian \cite{liu_variational_2021}:
\begin{equation}\label{poisson_hamiltonian}
 H=A^{\dagger}(I-|b\rangle\langle b|)A,
\end{equation}
where the prepared quantum state $|b\rangle\propto\mathbf{b}$ and $A$ is the coefficient matrix of the target linear system. The ground state $\ket{x}$ of Hamiltonian \eqref{poisson_hamiltonian} is proportional to the solution of the linear system \eqref{one_dimensional_linear_system} and \eqref{d_dimensional_linear_system} with $A=\widehat{A}$ and $A^{(d)}$, respectively. Here, we assume $n=2^{m}$ for some positive integer $m$. Additionally, we assume that there exists an efficient unitary operator $U_b$ that can prepare a quantum state $\ket{b}$ from the initial state $\ket{\overline{0}}$.

%  Here, it can be observed that $A^{\dagger}=A$ for the linear systems argued here and we always assume $A^{\dagger}=A$ below.

%we focus on the one-dimensional Poisson equation with the common boundary conditions of Neumann and Robin, and the mixed boundary conditions of Dirichlet, Neumann, and Robin and the d-dimensional Poisson equation with Dirichlet boundary conditions \cite{liu_variational_2021,Serigne2014_309,Serigne2014_372}.

\section{VARIATIONAL QUANTUM ALGORITHM FOR POISSON EQUATIONS BASED ON DECOMPOSITION OF SPARSE HAMILTONIANS}
\label{sec:VQA_SPARSE_HAMILTONIAN}
VQAs aim to find the ground state of the Hamiltonian \eqref{poisson_hamiltonian} by minimizing the loss function \cite{cerezo_variational_2021,herasymenko_diagrammatic_2021}:
\begin{equation}\label{VQA_loss_function}
\begin{aligned}
E(\vec{\theta})=\langle H \rangle = \langle \psi(\vec{\theta})|H|\psi(\vec{\theta})\rangle=\langle \psi(\vec{\theta})|A^2|\psi(\vec{\theta})\rangle-|\langle \mathbf{b}|A|\psi(\vec{\theta})\rangle|^2,
\end{aligned}
\end{equation}
where $\ket{\psi(\vec{\theta})}=U(\vec{\theta})\ket{+}^{\otimes N}$ with $\ket{+}=\frac{\ket{0}+\ket{1}}{2}$ is the variational wave-function ansatz used to approximate the ground state of the target Hamiltonian, $\vec{\theta}$ are the variational parameters to be optimized on the classical computer, and $N$ is the number of qubits which are required to encode the solutions for the argued linear systems. It is straightforward to see that $N$ is required to be $m$ and $md$, respectively, for the linear system \eqref{one_dimensional_linear_system} and \eqref{d_dimensional_linear_system}.

Efficient evaluation of the loss function $E(\vec{\theta})$ is limited due to the requirement that $A$ be decomposed into a sum of $O[\text{poly}(\text{log}_2n)]$ simple operators that can be easily measured or realized by quantum circuits.
Although the coefficient matrix $A$ of the discretized Poisson equation is three-sparse Hamiltonian, the graph with adjacency matrix $A$ is not bipartite, which implies that one can not directly decompose $A$ into a sum of Hermitian, one-sparse and self-inverse operators by using the idea introduced in \cite{berry_exponential_2014}. Therefore, we rewrite the loss function as
\begin{equation}
\begin{aligned}
E(\vec{\theta})=\langle H \rangle &= \langle \psi(\vec{\theta})|H|\psi(\vec{\theta})\rangle\\
&=\langle +|\langle \psi(\vec{\theta})|(\sigma_x\otimes A)^2|+\rangle|\psi(\vec{\theta})\rangle-|\langle+|\langle \mathbf{b}|\sigma_x\otimes A|+\rangle|\psi(\vec{\theta})\rangle|^2
\end{aligned}
\end{equation}
by noting that $\langle +|\sigma_x|+\rangle=1$ and $\langle +|\sigma^2_x|+\rangle=1$.
It is straightforward to see that the graph with adjacency matrix $\sigma_x\otimes A$ is bipartite. Hence, $\sigma_x\otimes A$ can be directly decomposed into a sum of Hermitian, one-sparse, and self-inverse operators.

Note that the decomposition of $(\sigma_x\otimes A)^2$ is natural when $\sigma_x\otimes A$ is decomposed into a sum of Hermitian, one-sparse, and self-inverse items since the product of such two decomposition items is still Hermitian, one-sparse, and self-inverse. Next, we first elaborate how to decompose $\sigma_x\otimes A$ into a sum of Hermitian, one-sparse, and self-inverse terms, then construct the quantum circuits to evaluate efficiently $E(\vec{\theta})$ for the linear systems \eqref{one_dimensional_linear_system} and \eqref{d_dimensional_linear_system} based on the method framework introduced in \cite{kirby_variational_2021}.

\subsection{EVALUATION OF $E(\vec{\theta})$ FOR THE LINEAR SYSTEM \eqref{one_dimensional_linear_system}}
\label{sec:VQA_one_dimention}
\begin{figure}
\begin{centering}
\includegraphics[scale=0.8]{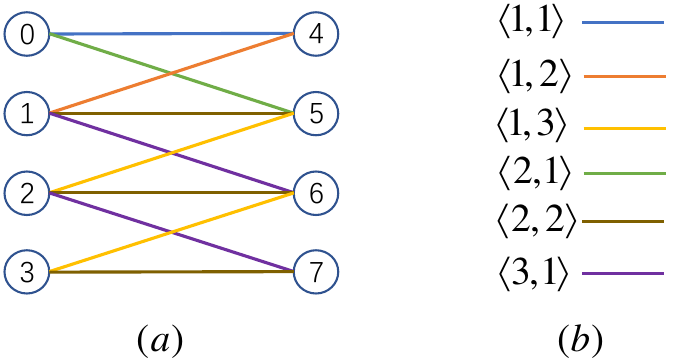}
\caption {Edge coloring decomposition of $\sigma_x\otimes\widehat{A}$ with $n=4$. ($a$) The bipartite graph with the adjacency matrix $\sigma_x\otimes\widehat{A}$. The edge $<u,v>$ formed by the nodes $u$ and $v$ is assigned to a color labeled by $<i,j>$ where $i$ denotes that the node $v$ is the $i$th neighbor of $u$ and $j$ denotes that $u$ is the $j$th neighbor of $v$. (b) The one-to-one correspondence between color and label $<i,j>$.}
\label{fig:decomposition_graph}
\end{centering}
\end{figure}

%Notice that $\widehat{A}$ and $\tilde{A}$ are $3$-sparse Hamiltonians.
For the linear system \eqref{one_dimensional_linear_system}, we first construct the bipartite graph corresponding to $\sigma_x\otimes \widehat{A}$ and adopt the edge coloring strategy in \cite{berry_exponential_2014} to obtain its decomposition into a sum of Hermitian and one-sparse terms. We take $n=4$ for example to illustrate the decomposition process. For $n=4$, $\sigma_x\otimes\widehat{A}$ has the following form:
\begin{equation}
\begin{aligned}
\sigma_x\otimes\widehat{A} &=\begin{array}{cc}
\mbox{}&
\begin{array}{cccccccc}\ 0\ \ \ &\ 1&\ \ \ 2&\ \ \ \ \,3\ &\ \ \ \ 4\ \ &\ \ \ 5\ \ &\ \ \ 6\ \ &\ \ \ 7 \end{array}\\
%\begin{array}{cccccccc}\text{ 0 }\text{ }&1\text{ }&2\text{ }&\text{ }3\text{ }&\text{ }4\text{ }&5\text{ }&6\text{ }&\text{ }7\text{ } \end{array}\\
\begin{array}{c}0\\1\\2\\3\\4\\5\\6\\7 \end{array}&
\left[\begin{array}{cccc|cccc}
\cline{5-8}
0&0&0&0&2-c&-1&0& \multicolumn{1}{c|}{0}\\
0&0&0&0&-1&2&-1&\multicolumn{1}{c|}{0}\\
0&0&0&0&0&-1&2&\multicolumn{1}{c|}{-1}\\
0&0&0&0&0&0&-1&\multicolumn{1}{c|}{2-d}\\
\cline{1-8}
\multicolumn{1}{|c}{2-c}&-1&0&0&0&0&0&0\\
\multicolumn{1}{|c}{-1}&2&-1&0&0&0&0&0\\
\multicolumn{1}{|c}{0}&-1&2&-1&0&0&0&0\\
\multicolumn{1}{|c}{0}&0&-1&2-d&0&0&0&0\\
\cline{1-4}
\end{array}\right],
\end{array}
\end{aligned}
\end{equation}
for which the corresponding bipartite graph is shown in Fig.~\ref{fig:decomposition_graph}($a$).

\begin{figure}
\begin{centering}
\includegraphics[scale=0.7]{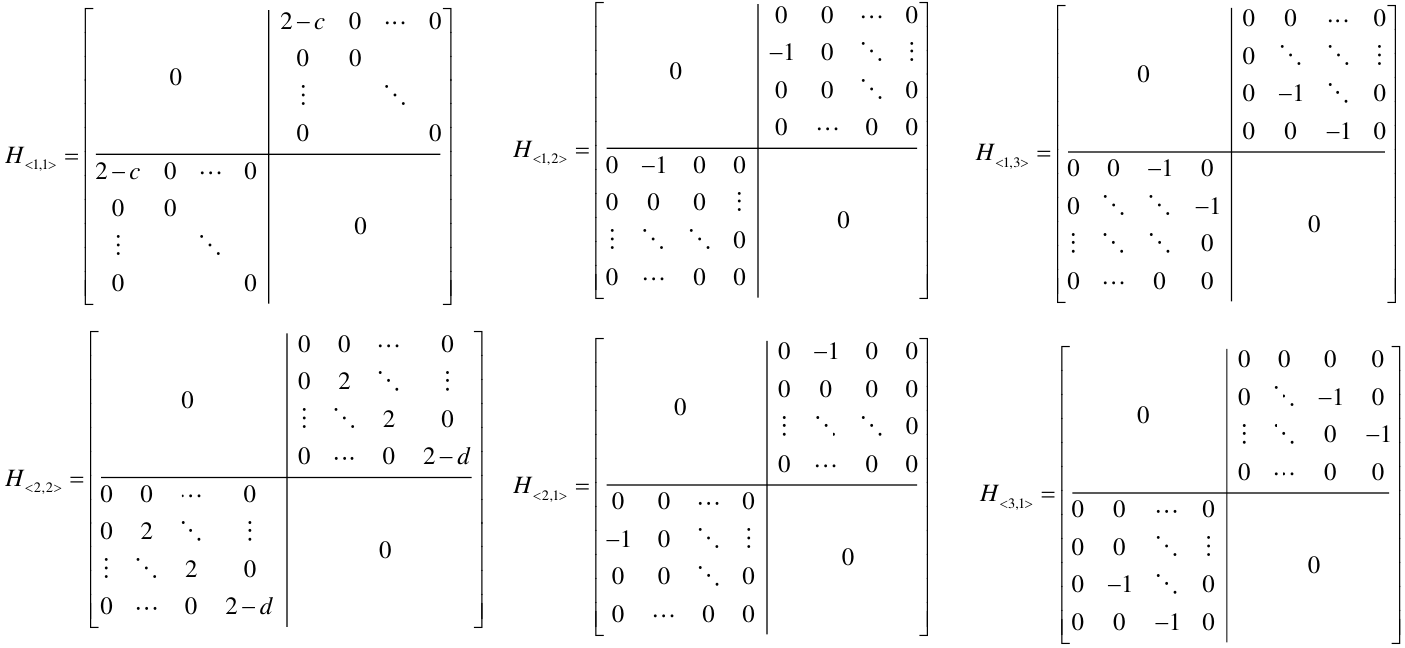}
\caption {The Hermitian and one-sparse decomposition items of $\sigma_x\otimes\widehat{A}$. $H_{<i,j>}$ corresponds to the adjacency matrix of the graph formed by the edges with the color labeled by $<i,j>$.}
\label{fig:decomposition_oneSparse}
\end{centering}
\end{figure}

Based on the coloring of the edges in the bipartite graph shown in Fig.~\ref{fig:decomposition_graph}($a$), $\sigma_x\otimes\widehat{A}$ can be decomposed into a sum of Hermitian and one-sparse items, which can be found in Fig.~\ref{fig:decomposition_oneSparse}. However, this decomposition process is not intuitive in obtaining the decomposition for the high-dimensional discretized Poisson equations, since the corresponding coefficient matrix $A^{(d)}$ will be more complex. Moreover, the quantum circuits for realizing the queries to the Hermitian, one-sparse, and self-inverse decomposition terms of $H_{<i,j>}$ are difficult to design due to the lack of regularity in the structure of $H_{<i,j>}$.

To tackle these difficulties, we note that $H_{<1,1>}+H_{<2,2>}$, $H_{<1,2>}+H_{<1,3>}$, and $H_{<2,1>}+H_{<3,1>}$ are all still Hermitian and one-sparse. Thus, $\sigma_x\otimes\widehat{A}$ can be first decomposed as
$\sigma_x\otimes\widehat{A}=H_{0}+H_{+1}+H_{-1}$, where
\begin{equation*}
\begin{aligned}
H_{0}=\footnotesize\left[
\begin{array}{cccc|cccc}
\cline{5-8}
0&0&0&0&2-c&0&0& \multicolumn{1}{c|}{0}\\
0&0&0&0&0&2&0&\multicolumn{1}{c|}{0}\\
0&0&0&0&0&0&2&\multicolumn{1}{c|}{0}\\
0&0&0&0&0&0&0&\multicolumn{1}{c|}{2-d}\\
\cline{1-8}
\multicolumn{1}{|c}{2-c}&0&0&0&0&0&0&0\\
\multicolumn{1}{|c}{0}&2&0&0&0&0&0&0\\
\multicolumn{1}{|c}{0}&0&2&0&0&0&0&0\\
\multicolumn{1}{|c}{0}&0&0&2-d&0&0&0&0\\
\cline{1-4}
 \end{array}
 \right],\,H_{+1}=\footnotesize\left[
\begin{array}{cccc|cccc}
\cline{5-8}
0&0&0&0&0&-1&0& \multicolumn{1}{c|}{0}\\
0&0&0&0&0&0&-1&\multicolumn{1}{c|}{0}\\
0&0&0&0&0&0&0&\multicolumn{1}{c|}{-1}\\
0&0&0&0&0&0&0&\multicolumn{1}{c|}{0}\\
\cline{1-8}
\multicolumn{1}{|c}{0}&0&0&0&0&0&0&0\\
\multicolumn{1}{|c}{-1}&0&0&0&0&0&0&0\\
\multicolumn{1}{|c}{0}&-1&0&0&0&0&0&0\\
\multicolumn{1}{|c}{0}&0&-1&0&0&0&0&0\\
\cline{1-4}
 \end{array}
 \right],\,H_{-1}=\footnotesize\left[
\begin{array}{cccc|cccc}
\cline{5-8}
0&0&0&0&0&0&0& \multicolumn{1}{c|}{0}\\
0&0&0&0&-1&0&0&\multicolumn{1}{c|}{0}\\
0&0&0&0&0&-1&0&\multicolumn{1}{c|}{0}\\
0&0&0&0&0&0&-1&\multicolumn{1}{c|}{0}\\
\cline{1-8}
\multicolumn{1}{|c}{0}&-1&0&0&0&0&0&0\\
\multicolumn{1}{|c}{0}&0&-1&0&0&0&0&0\\
\multicolumn{1}{|c}{0}&0&0&-1&0&0&0&0\\
\multicolumn{1}{|c}{0}&0&0&0&0&0&0&0\\
\cline{1-4}
 \end{array}
 \right],
\end{aligned}
\end{equation*}
from which we can see that we can directly obtain the decomposition of $\sigma_x\otimes\widehat{A}$ based on its structure instead of applying the edge coloring strategy. Moreover, the decomposition of $H_{0}$, $H_{+1}$, and $H_{-1}$ into a sum of Hermitian, one-sparse, and self-inverse operators can be obtained,  respectively, as
\begin{equation*}
\begin{aligned}
H_{0}&=2G_{0}-\frac{c}{2}(G_{0}-G_{0}^{f-})-\frac{d}{2}(G_{0}-G_{0}^{l-}),\\
H_{+1}&=-\frac{1}{2}(G_{+1}^{+}+G_{+1}^{-}),\\
H_{-1}&=-\frac{1}{2}(G_{-1}^{+}+G_{-1}^{-}),
\end{aligned}
\end{equation*}
where
\begin{equation*}
\begin{aligned}
G_{0}=\footnotesize\left[
\begin{array}{cccc|cccc}
\cline{5-8}
0&0&0&0&1&0&0& \multicolumn{1}{c|}{0}\\
0&0&0&0&0&1&0&\multicolumn{1}{c|}{0}\\
0&0&0&0&0&0&1&\multicolumn{1}{c|}{0}\\
0&0&0&0&0&0&0&\multicolumn{1}{c|}{1}\\
\cline{1-8}
\multicolumn{1}{|c}{1}&0&0&0&0&0&0&0\\
\multicolumn{1}{|c}{0}&1&0&0&0&0&0&0\\
\multicolumn{1}{|c}{0}&0&1&0&0&0&0&0\\
\multicolumn{1}{|c}{0}&0&0&1&0&0&0&0\\
\cline{1-4}
 \end{array}
 \right],\,G_{0}^{f-}=\footnotesize\left[
\begin{array}{cccc|cccc}
\cline{5-8}
0&0&0&0&-1&0&0& \multicolumn{1}{c|}{0}\\
0&0&0&0&0&1&0&\multicolumn{1}{c|}{0}\\
0&0&0&0&0&0&1&\multicolumn{1}{c|}{0}\\
0&0&0&0&0&0&0&\multicolumn{1}{c|}{1}\\
\cline{1-8}
\multicolumn{1}{|c}{-1}&0&0&0&0&0&0&0\\
\multicolumn{1}{|c}{0}&1&0&0&0&0&0&0\\
\multicolumn{1}{|c}{0}&0&1&0&0&0&0&0\\
\multicolumn{1}{|c}{0}&0&0&1&0&0&0&0\\
\cline{1-4}
 \end{array}
 \right],G_{0}^{l-}=\footnotesize\left[
\begin{array}{cccc|cccc}
\cline{5-8}
0&0&0&0&1&0&0& \multicolumn{1}{c|}{0}\\
0&0&0&0&0&1&0&\multicolumn{1}{c|}{0}\\
0&0&0&0&0&0&1&\multicolumn{1}{c|}{0}\\
0&0&0&0&0&0&0&\multicolumn{1}{c|}{-1}\\
\cline{1-8}
\multicolumn{1}{|c}{1}&0&0&0&0&0&0&0\\
\multicolumn{1}{|c}{0}&1&0&0&0&0&0&0\\
\multicolumn{1}{|c}{0}&0&1&0&0&0&0&0\\
\multicolumn{1}{|c}{0}&0&0&-1&0&0&0&0\\
\cline{1-4}
 \end{array}
 \right]
\end{aligned}
\end{equation*}
and
\begin{equation*}
\begin{aligned}
G_{+1}^{\pm}=\footnotesize\left[
\begin{array}{cccc|cccc}
\cline{5-8}
0&0&0&0&0&1&0& \multicolumn{1}{c|}{0}\\
0&0&0&0&0&0&1&\multicolumn{1}{c|}{0}\\
0&0&0&0&0&0&0&\multicolumn{1}{c|}{1}\\
0&0&0&\pm1&0&0&0&\multicolumn{1}{c|}{0}\\
\cline{1-8}
\multicolumn{1}{|c}{0}&0&0&0&\pm1&0&0&0\\
\multicolumn{1}{|c}{1}&0&0&0&0&0&0&0\\
\multicolumn{1}{|c}{0}&1&0&0&0&0&0&0\\
\multicolumn{1}{|c}{0}&0&1&0&0&0&0&0\\
\cline{1-4}
 \end{array}
 \right],G_{-1}^{\pm}=\footnotesize\left[
\begin{array}{cccc|cccc}
\cline{5-8}
\pm1&0&0&0&0&0&0& \multicolumn{1}{c|}{0}\\
0&0&0&0&1&0&0&\multicolumn{1}{c|}{0}\\
0&0&0&0&0&1&0&\multicolumn{1}{c|}{0}\\
0&0&0&0&0&0&1&\multicolumn{1}{c|}{0}\\
\cline{1-8}
\multicolumn{1}{|c}{0}&1&0&0&0&0&0&0\\
\multicolumn{1}{|c}{0}&0&1&0&0&0&0&0\\
\multicolumn{1}{|c}{0}&0&0&1&0&0&0&0\\
\multicolumn{1}{|c}{0}&0&0&0&0&0&0&\pm1\\
\cline{1-4}
 \end{array}
 \right].
\end{aligned}
\end{equation*}
It is straightforward to verity that $G_{0}$, $G_{0}^{f-}$, $G_{0}^{l-}$, $G_{+1}^{\pm}$, and $G_{-1}^{\pm}$ are all Hermitian, one-sparse and self-inverse operators.

Consequently, the decomposition of $\sigma_x\otimes \widehat{A}$ into a sum of Hermitian, one-sparse, and self-inverse operators is obtained for general $n$ as
\begin{equation}
\begin{aligned}\label{one_A_decomposition}
\sigma_x\otimes \widehat{A} =& 2G_{0} - \frac{1}{2}(G_{+1}^{+}+G_{+1}^{-})-\frac{1}{2}(G_{-1}^{+}+G_{-1}^{-})
-\frac{c}{2}(G_0-G_0^{f-})-\frac{d}{2}(G_0-G_0^{l-}),\\
=&\left(2-\frac{c+d}{2}\right)G_0+\frac{c}{2}G_0^{f-}+\frac{d}{2}G_0^{l-}
-\frac{1}{2}(G_{+1}^{+}+G_{+1}^{-}+G_{-1}^{+}+G_{-1}^{-}).
\end{aligned}
\end{equation}

On the other hand, although the decomposition of $(\sigma_x\otimes \widehat{A})^2$ is obtained based on Eq.~\eqref{one_A_decomposition}, $\sigma_x\otimes \widehat{A}^2$ can be directly decomposed into a sum of Hermitian, one-sparse and self-inverse operators with a much smaller number of terms since $\widehat{A}^2$ is a four-sparse Hamiltonian of the form
\begin{equation}
\begin{aligned}
\widehat{A}^2&= \left[
\begin{array}{cccccc}
6 & -4 & 1& &  &0 \\
-4 & 6 & -4 & 1 &  & \\
1 & \ddots& \ddots & \ddots& \ddots & \\
 & \ddots & \ddots &\ddots & \ddots & 1\\
 &  & 1&-4 & 6 & -4 \\
0&   &  &1 & -4 & 6 \\
 \end{array}
 \right]+\left[
\begin{array}{cccccc}
0 & c &  & &  &  \\
c & 0 & 0 &  &  & \\
  & \ddots & \ddots & \ddots&  & \\
 &  & 0 &0 & 0 & \\
 &  & &0 & 0 & d \\
 &   &  &  & d & 0 \\
 \end{array}
 \right]-\left[
\begin{array}{cccccc}
4c+1-c^2&  & & & & 0 \\
 & 0& & &  & \\
 &  & \ddots & & & \\
 &  & & 0 & &  \\
 &  & & & 0 & \\
0&  & & & & 4d+1-d^2\\
 \end{array}
 \right].
 \end{aligned}
\end{equation}
Due to its structure, the decomposition of $\sigma_x\otimes\widehat{A}^2$ is obtained as
\begin{align}\label{A2_decomposition}
\sigma_x\otimes \widehat{A}^2 = &6G_0-2(G_{+1}^{+}+G_{+1}^{-}+G_{-1}^{+}+G_{-1}^{-})
+\frac{1}{2}(G_{+2}^{+}+G_{+2}^{-}+G_{-2}^{+}+G_{-2}^{-})\nonumber\\
&+\frac{c}{2}(G_{1}^{f+}-G_{1}^{f-})+\frac{d}{2}(G_{1}^{l+}-G_{1}^{l-})
-\frac{4c+1-c^2}{2}(G_0-G_0^{f-})-\frac{4d+1-d^2}{2}(G_0-G_0^{l-})\nonumber\\
=&\left(5+\frac{c^2+d^2}{2}-2c-2d\right)G_0+\frac{4c+1-c^2}{2}G_0^{f-}
+\frac{4d+1-d^2}{2}G_0^{l-}+\frac{c}{2}(G_{1}^{f+}-G_{1}^{f-})
+\frac{d}{2}(G_{1}^{l+}-G_{1}^{l-})\nonumber\\
&-2(G_{+1}^{+}+G_{+1}^{-}+G_{-1}^{+}+G_{-1}^{-})+\frac{1}{2}
(G_{+2}^{+}+G_{+2}^{-}+G_{-2}^{+}+G_{-2}^{-}),
 \end{align}
where the decomposition items will be exactly illustrated later. Therefore, to evaluate the loss function more efficiently, we compute the value $\langle+|\langle \psi(\vec{\theta})|\sigma_x\otimes \widehat{A}^2|+\rangle|\psi(\vec{\theta})\rangle$ instead of $\langle+|\langle \psi(\vec{\theta})|(\sigma_x\otimes \widehat{A})^2|+\rangle|\psi(\vec{\theta})\rangle$.

Let $|q_m\cdots q_0\rangle$, or $|g\rangle$ with $g=\sum_{i=0}^{i=m}q_i2^{i}$,  be a computational basis state with $m+1$ qubits. In the computational basis the decomposition terms can be expressed as
\begin{align}\label{one_dimension_decomposition_effects}
&G_0:|q_m\cdots q_0\rangle \rightarrow |(1-q_m)q_{m-1}\cdots q_0\rangle,\nonumber\\
&G_0^{f-}:|q_m\cdots q_0\rangle \rightarrow (-1)^{\prod_{i=0}^{m-1}(1-q_i)}|(1-q_m)q_{m-1}\cdots q_0\rangle,\nonumber\\
&G_0^{l-}:|q_m\cdots q_0\rangle \rightarrow (-1)^{\prod_{i=0}^{m-1}q_i}|(1-q_m)q_{m-1}\cdots q_0\rangle,\nonumber\\
&G_{+1}^{\pm}:|g\rangle \rightarrow \pm \prod_{i=0}^{m-1} q_i|g\rangle+\left(1-\prod_{i=0}^{m-1} q_i\right)|g+2^m+1\rangle,~g=0,1,\cdots,2^m-1,\nonumber\\
&\qquad \ \ |g\rangle \rightarrow \pm \prod_{i=0}^{m-1} (1-q_i)|g\rangle+\left(1-\prod_{i=0}^{m-1} (1-q_i)\right)|g-2^m-1\rangle, ~g=2^{m},\cdots,2^{m+1}-1,\nonumber\\
&G_{-1}^{\pm}:|g\rangle \rightarrow \pm \prod_{i=0}^{m-1} (1-q_i)|g\rangle+\left(1-\prod_{i=0}^{m-1} (1-q_i)\right)|g+2^m-1\rangle, ~g=0,1,\cdots,2^{m}-1,\nonumber\\
&\qquad \ \ |g\rangle \rightarrow \pm \prod_{i=0}^{m-1} q_i|g\rangle+\left(1-\prod_{i=0}^{m-1} q_i\right)|g-2^m+1\rangle,g=2^m,\cdots,2^{m+1}-1,\nonumber\\
&G_{1}^{f\pm}:|q_m\cdots q_0\rangle \rightarrow \pm \prod_{i=1}^{m-1}(1-q_i)|(1-q_m)q_{m-1}\cdots q_1(1-q_0)\rangle+\left(1-\prod_{i=1}^{m-1}(1-q_i)\right)|q_m\cdots q_0\rangle,\nonumber\\
&G_{1}^{l\pm}:|q_m\cdots q_0\rangle \rightarrow \pm \prod_{i=1}^{m-1}q_i|(1-q_m)q_{m-1}\cdots q_1(1-q_0)\rangle+(1-\prod_{i=1}^{m-1}q_i)|q_m\cdots q_0\rangle,\nonumber\\
&G_{+2}^{\pm}:|g\rangle \rightarrow |g+2^m+2\rangle,\,\,g=0,1,\cdots,2^m-3,\nonumber\\
&\qquad \ \ |g\rangle \rightarrow \pm |g\rangle,\,\,g=2^m-2,2^m-1,2^m,2^m+1,\nonumber\\
&\qquad \ \ |g\rangle \rightarrow |g-2^m-2\rangle,\,\,g=2^m+2,\cdots,2^{m+1}-1,\nonumber\\
&G_{-2}^{\pm}:|g\rangle \rightarrow \pm |g\rangle,\,\,g=0,1,2^{m+1}-2,2^{m+1}-1,\nonumber\\
&\qquad \ \ |g\rangle \rightarrow |g+2^m-2\rangle,\,\,g=2,\cdots,2^m-1,\nonumber\\
&\qquad \ \ |g\rangle \rightarrow |g-2^m+2\rangle,\,\,g=2^m,\cdots,2^{m+1}-3,
\end{align}
which can be easily obtained from the decomposition term structure of the example analyzed above.
Sequentially, we have
\begin{align}
\langle \mathbf{b}|A|\psi(\vec{\theta})\rangle&=\langle+|\langle \mathbf{b}|\sigma_x\otimes A|+\rangle|\psi(\vec{\theta})\rangle\nonumber\\
&=\left(2-\frac{c+d}{2}\right)\langle+|\langle \mathbf{b}|G_0|+\rangle|\psi(\vec{\theta})\rangle+\frac{c}{2}\langle+|\langle \mathbf{b}| G_0^{f-}|+\rangle|\psi(\vec{\theta})\rangle+\frac{d}{2}\langle+|\langle \mathbf{b}| G_0^{l-}|+\rangle|\psi(\vec{\theta})\rangle\nonumber\\
&-\frac{1}{2}\sum_{G\in\{G_{+1}^{+},G_{+1}^{-},G_{-1}^{+},G_{-1}^{-}\}}\langle+|\langle \mathbf{b}|G|+\rangle|\psi(\vec{\theta})\rangle
\end{align}
and
\begin{align}
\langle \psi(\vec{\theta})|A^2|\psi(\vec{\theta})\rangle &=\langle +|\langle \psi(\vec{\theta})|\sigma_x\otimes A^2|+\rangle|\psi(\vec{\theta})\rangle\nonumber\\
&=\left(5+\frac{c^2+d^2}{2}-2c-2d\right)\langle +|\langle\psi(\vec{\theta})|G_0|+\rangle|\psi(\vec{\theta})\rangle+\frac{4c+1-c^2}{2}\langle +|\langle \psi(\vec{\theta})|G_0^{f-}|+\rangle|\psi(\vec{\theta})\rangle\nonumber\\
&+\frac{4d+1-d^2}{2}\langle +|\langle \psi(\vec{\theta})|G_0^{l-}|+\rangle|\psi(\vec{\theta})\rangle+\frac{c}{2}\left(\langle +|\langle \psi(\vec{\theta})|G_{1}^{f+}|+\rangle|\psi(\vec{\theta})\rangle-\langle +|\langle \psi(\vec{\theta})|G_{1}^{f-}|+\rangle|\psi(\vec{\theta})\rangle\right)\nonumber\\
&+\frac{d}{2}\left(\langle +|\langle \psi(\vec{\theta})|G_{1}^{l+}|+\rangle|\psi(\vec{\theta})\rangle-\langle +|\langle \psi(\vec{\theta})|G_{1}^{l-}|+\rangle|\psi(\vec{\theta})\rangle\right)-2\sum_{G\in\{G_{+1}^{+},G_{+1}^{-},G_{-1}^{+},G_{-1}^{-}\}}\langle +|\langle \psi(\vec{\theta})|G|+\rangle|\psi(\vec{\theta})\rangle\nonumber\\
&+\frac{1}{2}\sum_{G\in\{G_{+2}^{+},G_{+2}^{-},G_{-2}^{+},G_{-2}^{-}\}}\langle +|\langle \psi(\vec{\theta})|G|+\rangle|\psi(\vec{\theta})\rangle.
\end{align}

Let $G$ be one of the Hermitian, one-sparse and self-inverse terms. The problem of evaluating $E(\vec{\theta})$ for $A=\widehat{A}$ can be transformed to the problem of computing the values of the terms $\langle +|\langle \psi(\vec{\theta})|G|+\rangle|\psi(\vec{\theta})\rangle$ and $\langle+|\langle \mathbf{b}|G|+\rangle|\psi(\vec{\theta})\rangle$.
Note that the total number of items $\langle+|\langle \mathbf{b}|G|+\rangle|\psi(\vec{\theta})\rangle$ and $\langle +|\langle \psi(\vec{\theta})|G|+\rangle|\psi(\vec{\theta})\rangle$ which are required to be computed are at most 7 and 15, respectively. Particularly, only 5 terms $\langle+|\langle \mathbf{b}|G|+\rangle|\psi(\vec{\theta})\rangle$ and 11 terms $\langle +|\langle \psi(\vec{\theta})|G|+\rangle|\psi(\vec{\theta})\rangle$ are needed to be computed for the one-dimensional Poisson equation with Dirichlet boundary conditions. We emphasize that the total number of the items is independent of $n$.

\begin{figure}
\begin{centering}
\includegraphics[scale=0.4]{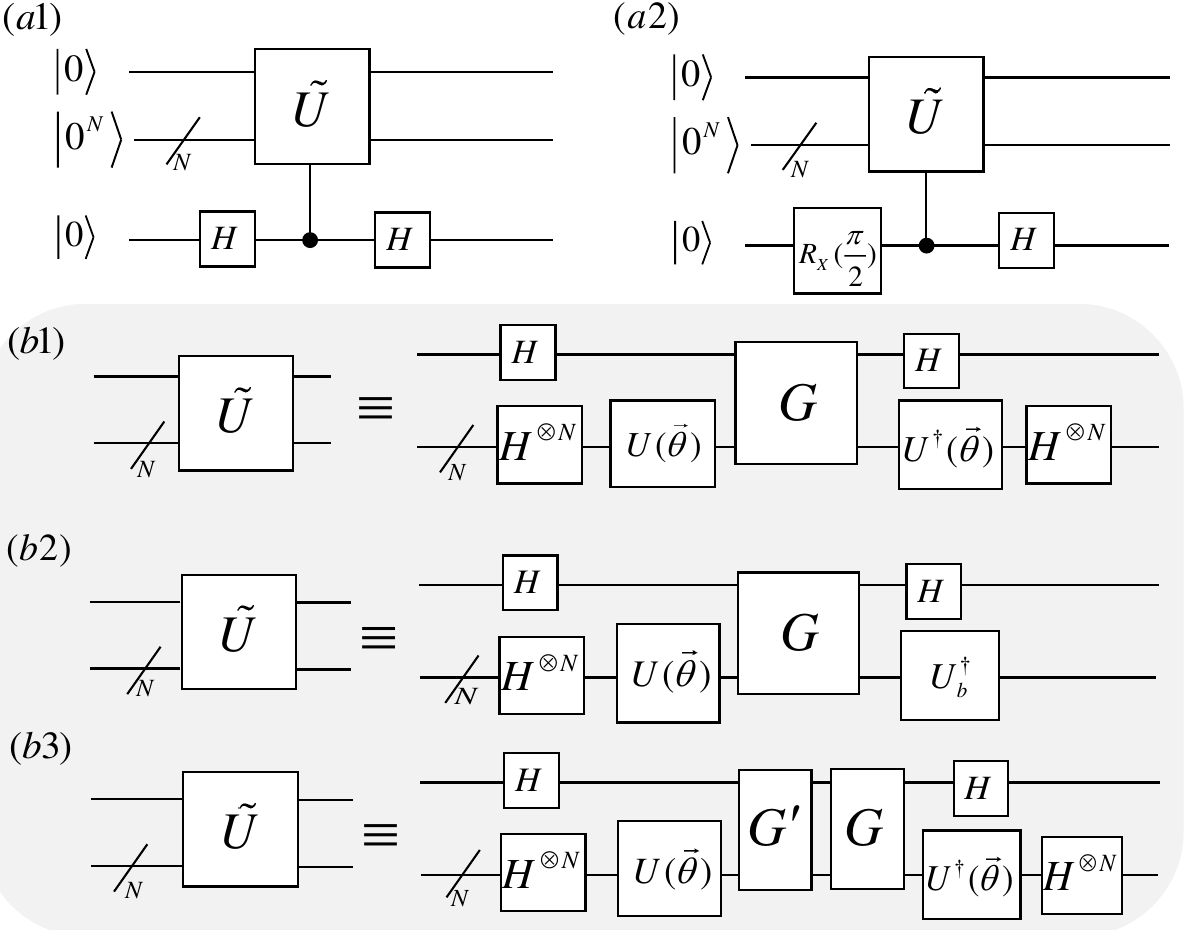}
\caption {Hadamard test circuits for estimating the values of the items $\langle +|\langle \psi(\vec{\theta})|G|+\rangle|\psi(\vec{\theta})\rangle$, $\langle+|\langle \mathbf{b}|G|+\rangle|\psi(\vec{\theta})\rangle$ and $\langle +|\langle \psi(\vec{\theta})|GG'|+\rangle|\psi(\vec{\theta})\rangle$.
($a$1) The Hadamard test circuit to estimate the real part of the items $\langle +|\langle \psi(\vec{\theta})|G|+\rangle|\psi(\vec{\theta})\rangle$, $\langle+|\langle \mathbf{b}|G|+\rangle|\psi(\vec{\theta})\rangle$ or $\langle +|\langle \psi(\vec{\theta})|GG'|+\rangle|\psi(\vec{\theta})\rangle$ when the controlled unitary operator is selected to be of the forms shown in (b1), (b2) or (b3), respectively.
($a$2)  The Hadamard test circuit to estimate the imaginary part of the items $\langle +|\langle \psi(\vec{\theta})|G|+\rangle|\psi(\vec{\theta})\rangle$, $\langle+|\langle \mathbf{b}|G|+\rangle|\psi(\vec{\theta})\rangle$ or $\langle +|\langle \psi(\vec{\theta})|GG'|+\rangle|\psi(\vec{\theta})\rangle$ when the controlled unitary operator is selected to be the form shown in (b1), (b2) or (b3), respectively.
(b1)-(b3) The circuits for the controlled unitary operator used in estimating the value of $\langle +|\langle \psi(\vec{\theta})|G|+\rangle|\psi(\vec{\theta})\rangle$, $\langle+|\langle \mathbf{b}|G|+\rangle|\psi(\vec{\theta})\rangle$ and $\langle +|\langle \psi(\vec{\theta})|GG'|+\rangle|\psi(\vec{\theta})\rangle$.}
\label{fig:circuit_VQA}
\end{centering}
\end{figure}

Here, we design the corresponding quantum circuits for the decomposition terms on the right-hand side of \eqref{one_A_decomposition} and \eqref{A2_decomposition} (see Appendix~\ref{sec:appendix_A}). We adopt the Hadamard test circuit to obtain the values of the terms $\langle +|\langle \psi(\vec{\theta})|G|+\rangle|\psi(\vec{\theta})\rangle$ and $\langle+|\langle \mathbf{b}|G|+\rangle|\psi(\vec{\theta})\rangle$.
Specifically, after $M$ repetitions of the circuits shown in Figure~\ref{fig:circuit_VQA}($a$1) and Figure~\ref{fig:circuit_VQA}($a$2) with controlled unitary operator $\tilde{U}$, one obtains the probabilities of observing zero on the ancilla qubit, denoted as $P_{R}$ and $P_{I}$, respectively. The value of the item $\langle0|\tilde{U}|0\rangle$ can be evaluated as
\begin{equation}\label{value_hadamand_test}
\begin{aligned}
\langle0|\tilde{U}|0\rangle=(2P_{R}-1)+i(2P_{I}-1).
\end{aligned}
\end{equation}
Therefore the values of the items $\langle +|\langle \psi(\vec{\theta})|G|+\rangle|\psi(\vec{\theta})\rangle$ and $\langle+|\langle \mathbf{b}|G|+\rangle|\psi(\vec{\theta})\rangle$ can be estimated by using the Eq.~\eqref{value_hadamand_test} when the operator $\tilde{U}$ is selected to be of the forms shown in Fig.~\ref{fig:circuit_VQA}(b1) and Fig.~\ref{fig:circuit_VQA}(b2), respectively.
%  shown in Figure~\ref{fig:circuit_VQA}(b1), respectively, one obtains the probabilities of observing zero on the ancilla qubit, denoted as $P_{R}$ and $P_{I}$. The value of the term $\langle +|\langle \psi(\vec{\theta})|G|+\rangle|\psi(\vec{\theta})\rangle$ can be obtained as
%\begin{equation}\label{G_expect_psi_psi}
%\begin{aligned}
%\langle +|\langle \psi(\vec{\theta})|G|+\rangle|\psi(\vec{\theta})\rangle=(2P_{R}-1)+i(2P_{I}-1).
% \end{aligned}
%\end{equation}
%Similarly, we have
%\begin{equation}\label{G_expect_b_psi}
%\begin{aligned}
%\langle+|\langle \mathbf{b}|G|+\rangle|\psi(\vec{\theta})\rangle=(2Q_{R}-1)+i(2Q_{I}-1),
% \end{aligned}
%\end{equation}
%where $Q_{R}$ and $Q_{I}$ corresponds to the probabilities of observing zero on the ancilla qubit after executing $M$ circuits shown in Figure~\ref{fig:circuit_VQA}(a1) and Figure~\ref{fig:circuit_VQA}(a2) with controlled unitary operator shown in Figure~\ref{fig:circuit_VQA}(b2), respectively.

\subsection{EVALUATION OF $E(\vec{\theta})$ FOR THE LINEAR SYSTEM \eqref{d_dimensional_linear_system}}
\label{sec:VQA_d_dimention}
With respect to the linear system \eqref{d_dimensional_linear_system} with $d\ge2$, we have
\begin{equation}\label{A_d_equation}
\begin{aligned}
A^{(d)}=\sum_{s=0,s+t=d-1}^{d-1}I_{s}\otimes \tilde{A}\otimes I_{t},
\end{aligned}
\end{equation}
with $I_{x} = \underbrace{I \otimes \cdots \otimes I}_{x}$.
The term $I_{s}\otimes \tilde{A}\otimes I_{t}$ is shown to be of the form
\begin{widetext}
\begin{equation}
\begin{aligned}
 I_{s}\otimes \tilde{A}\otimes I_{t} &=\small\begin{blockarray}{ccccccccccccccccc}
\begin{block}{[cccccccccccccccc]c}
\textbf{[2]} & \textbf{[-1]} & & & & & & & & & & & & & & &\overbrace{0\cdots00}^{ms}\overbrace{0\cdots00}^{m}\\
\textbf{[-1]}& \textbf{[2]} &\textbf{[-1]} &  & & & & & & & & & & & & &0\cdots000\cdots01 \\
&\ddots & \ddots &\ddots &  & & & & & & & & & & &  &\vdots\\
&  & \ddots & \ddots & \textbf{[-1]} & & & & & & & & & & & &0\cdots001\cdots10 \\
& &  & \textbf{[-1]} &\textbf{[2]}& & & & & & & & & & & &0\cdots001\cdots11\\
\BAhline& & & & & \textbf{[2]} &  \textbf{[-1]} & & & & & & & & & &0\cdots010\cdots00\\
& & & & & \textbf{[-1]} &\textbf{[2]} & \textbf{[-1]}  & & & & & & & & &0\cdots010\cdots01\\
& & & & & & \ddots &\ddots & \ddots& & & & & & & &\vdots\\
& & & & & & &\ddots & \ddots&\textbf{[-1]} & & & & & & &0\cdots011\cdots10\\
& & & & & & & &\textbf{[-1]} & \textbf{[2]} & & & & & & &0\cdots011\cdots11\\
\BAhline& & & & & & & & & &\ddots& & & & & &\vdots\\
\BAhline& & & & & & & & & & &\textbf{[2]} & \textbf{[-1]} & & & & 1\cdots110\cdots00\\
& & & & & & & & & & &\textbf{[-1]}& \textbf{[2]} &\textbf{[-1]} &  & & 1\cdots110\cdots01\\
& & & & & & & & & & & &\ddots & \ddots &\ddots &  &\vdots\\
& & & & & & & & & & & &  & \ddots & \ddots & \textbf{[-1]} & 1\cdots111\cdots10\\
& & & & & & & & & & & & &  & \textbf{[-1]} &\textbf{[2]} &1\cdots111\cdots11\\
\end{block}
\end{blockarray}\\
\end{aligned}
\end{equation}
\end{widetext}
with
\begin{equation}
\begin{aligned}
\textbf{[c]}&=\begin{blockarray}{ccccc}
\begin{block}{[cccc]c}
c & & & &\overbrace{0\cdots00}^{mt}\\
 & c & & &0\cdots01\\
& & \ddots & &\vdots\\
& & & c & 1\cdots11\\
\end{block}
\end{blockarray}.
\end{aligned}
\end{equation}

Therefore, it is observed that $I_{s}\otimes \tilde{A}\otimes I_{t}$ is a three-sparse Hamiltonian. Consequently, $\sigma_x\otimes I_{s}\otimes \tilde{A}\otimes I_{t}$ can be decomposed into a sum of Hermitian, one-sparse and self-inverse terms:
\begin{equation}\label{A_d_one_term_decomposition}
\begin{aligned}
\sigma_x\otimes I_{s} \otimes \tilde{A}\otimes I_{t} =& 2G_{d,0} - \frac{1}{2}\left(G_{d,+2^{mt}}^{+}+G_{d,+2^{mt}}^{-}\right)-\frac{1}{2}\left(G_{d,-2^{mt}}^{+}+G_{d,-2^{mt}}^{-}\right),
 \end{aligned}
\end{equation}
where $s=0,1,\cdots,d-1$ and $t=d-1-s$. Let $|q_{md}\cdots q_{0}\rangle$, or $|g\rangle$ with $g=\sum_{i=0}^{i=md}q_i2^{i}$, be a computational basis state with $md+1$ qubits. In the computational basis we have
\begin{align}\label{d_dimension_decomposition_effects}
&G_{d,0}:|q_{md}\cdots q_{0}\rangle \rightarrow |(1-q_{md})q_{md-1}\cdots q_{0}\rangle,\nonumber\\
&G_{d,+2^{mt}}^{\pm}:|g\rangle \rightarrow \pm \prod_{i=0}^{i=m-1}q_{mt+i}|g\rangle+(1-\prod_{i=0}^{i=m-1}q_{mt+i})|g+2^{md}+2^{mt}\rangle,g=0,\cdots,2^{md}-1,\nonumber\\
&\qquad \quad \ \ \ |g\rangle \rightarrow \pm \prod_{i=0}^{i=m-1}(1-q_{mt+i})|g\rangle+\left(1-\prod_{i=0}^{i=m-1}(1-q_{mt+i})\right)|g-2^{md}-2^{mt}\rangle ,g=2^{md},\cdots,2^{md+1}-1,\nonumber\\
&G_{d,-2^{mt}}^{\pm}:|g\rangle \rightarrow \pm \prod_{i=0}^{i=m-1}(1-q_{mt+i})|g\rangle+\left(1-\prod_{i=0}^{i=m-1}(1-q_{mt+i})\right)|g+2^{md}-2^{mt}\rangle, g=0,\cdots,2^{md}-1,\nonumber\\
&\qquad \quad \ \ \  |g\rangle \rightarrow \pm \prod_{i=0}^{i=m-1}q_{mt+i}|g\rangle +(1-\prod_{i=0}^{i=m-1}q_{mt+i})|g-2^{md}+2^{mt}\rangle, g=2^{md},\cdots,2^{md+1}-1.
\end{align}
Accordingly, the quantum circuits for the decomposition terms on the right-hand side of \eqref{A_d_one_term_decomposition} are designed (see Appendix~\ref{sec:appendix_A}).

Combining Eq.~\eqref{A_d_equation} with the decomposition \eqref{A_d_one_term_decomposition}, we obtain
\begin{equation}\label{A_d_term_decomposition}
\begin{aligned}
\sigma_x \otimes A^{(d)}&=2dG_{d,0}-\frac{1}{2}\sum_{t=0}^{d-1}\Big(G_{d,+2^{mt}}^{+}+G_{d,+2^{mt}}^{-}
+G_{d,-2^{mt}}^{+}+G_{d,-2^{mt}}^{-}\Big),\\
\langle \mathbf{b}|A^{(d)}|\psi(\vec{\theta})\rangle&=\langle+|\langle \mathbf{b}|\sigma_x\otimes A^{(d)}|+\rangle|\psi(\vec{\theta})\rangle\\
&=2d\langle+|\langle \mathbf{b}|G_{d,0}|+\rangle|\psi(\vec{\theta})\rangle-\frac{1}{2}\sum_{t=0}^{d-1}
\Big[\langle+|\langle \mathbf{b}|G_{d,+2^{mt}}^{+}|+\rangle|\psi(\vec{\theta})\rangle+\langle+|\langle \mathbf{b}|G_{d,+2^{mt}}^{-}|+\rangle|\psi(\vec{\theta})\\
&+\langle+|\langle \mathbf{b}|G_{d,-2^{mt}}^{+}|+\rangle|\psi(\vec{\theta})\rangle+\langle+|\langle \mathbf{b}|G_{d,-2^{mt}}^{-}|+\rangle|\psi(\vec{\theta})\rangle\Big].
 \end{aligned}
\end{equation}
In other words, the total number of the items required to be computed to obtain the value of $\langle \mathbf{b}|A^{(d)}|\psi(\vec{\theta})\rangle$ is $4d+1$, which is also independent of $n$. Additionally, the value of the items on the right-hand side of Eq.~\eqref{A_d_term_decomposition} can be obtained by using the Eq.~\eqref{value_hadamand_test} when the controlled operator $\tilde{U}$ is selected to be of the form shown in Fig.~\ref{fig:circuit_VQA}(b2).

%Based on the decomposition \eqref{A_d_term_decomposition}, $\langle+|\langle \mathbf{b}|\sigma_x\otimes A^{(b)}|+\rangle|\psi(\vec{\theta})\rangle$ can be computed as
%Similar to the discussion in Sec.~\ref{sec:VQA_one_dimention}, \eqref{G_expect_b_psi} can also be used to evaluate the values of the terms on the right hand of \eqref{A_d_dimention_expect}.

On the other hand, $(\sigma_x \otimes A^{(d)})^2$ can be expressed as a linear combination of $(4d+1)^2-(4d+1)$ Hermitian, one-sparse and self-inverse operators, for which the identity operator is ignored.
%We can regard the number of decomposition terms for $\sigma^2_x \otimes (A^{(d)})^2$ as $(4d+1)^2-(4d+1)$, since
Thus, we have
\begin{align}\label{A_d_2_expect}
\langle \psi(\vec{\theta})|(A^{(d)})^2|\psi(\vec{\theta})\rangle&=\langle +|\langle \psi(\vec{\theta})|(\sigma_x \otimes A^{(d)})^2|+\rangle|\psi(\vec{\theta})\rangle\nonumber\\
&=(4d^2+d)-d\sum_{t=0}^{d-1}\Big[\langle +|\langle \psi(\vec{\theta})|G_{d,0}G_{d,+2^{mt}}^{+}|+\rangle|\psi(\vec{\theta})\rangle+\langle +|\langle \psi(\vec{\theta})|G_{d,0}G_{d,+2^{mt}}^{-}|+\rangle|\psi(\vec{\theta})\rangle\nonumber\\
&+\langle +|\langle \psi(\vec{\theta})|G_{d,0}G_{d,-2^{mt}}^{+}|+\rangle|\psi(\vec{\theta})\rangle+\langle +|\langle \psi(\vec{\theta})|G_{d,0}G_{d,-2^{mt}}^{-}|+\rangle|\psi(\vec{\theta})\rangle\Big]\nonumber\\
&-d\sum_{t=0}^{d-1}\Big[\langle +|\langle \psi(\vec{\theta})|G_{d,+2^{mt}}^{+}G_{d,0}|+\rangle|\psi(\vec{\theta})\rangle+\langle +|\langle \psi(\vec{\theta})|G_{d,+2^{mt}}^{-}G_{d,0}|+\rangle|\psi(\vec{\theta})\rangle\nonumber\\
&+\langle +|\langle \psi(\vec{\theta})|G_{d,-2^{mt}}^{+}G_{d,0}|+\rangle|\psi(\vec{\theta})\rangle+\langle +|\langle \psi(\vec{\theta})|G_{d,-2^{mt}}^{-}G_{d,0}|+\rangle|\psi(\vec{\theta})\rangle\Big]\nonumber\\
&+\frac{1}{4}\sum_{G,G'\in S,G'\neq G}\Big[\langle +|\langle \psi(\vec{\theta})|GG'|+\rangle|\psi(\vec{\theta})\rangle+\langle +|\langle \psi(\vec{\theta})|G'G|+\rangle|\psi(\vec{\theta})\rangle\Big],
\end{align}
where $S=\cup_{t=0}^{d-1}\left\{G_{d,+2^{mt}}^{+}, G_{d,+2^{mt}}^{-},G_{d,-2^{mt}}^{+},G_{d,-2^{mt}}^{-}\right\}$. Thus, it is seen that only $(4d+1)^2-(4d+1)$ items
$\langle +|\langle \psi(\vec{\theta})|GG'|+\rangle|\psi(\vec{\theta})\rangle$
are needed to be computed in order to obtain the value of  $\langle \psi(\vec{\theta})|(A^{(d)})^2|\psi(\vec{\theta})\rangle$, where $G$ and $G'$ denote two different Hermitian, one-sparse and self-inverse operators. Moreover, the value of the item $\langle +|\langle \psi(\vec{\theta})|GG'|+\rangle|\psi(\vec{\theta})\rangle$ can be evaluated by using Eq.~\eqref{value_hadamand_test} when the controlled operator $\tilde{U}$ is selected to be of the form shown in Fig.~\ref{fig:circuit_VQA}(b3).

\section{NUMERICAL EXPERIMENTS}
\label{sec:Experiments}
We test our algorithm for the one-dimensional Poisson equations with Dirichlet boundary conditions. For convenience, we choose $|b\rangle=\sum_{i=0}^{m-1}(\frac{|0\rangle+|1\rangle}{\sqrt{2}})^{\otimes m}$. Namely, the operator $U_b$ in Fig.~\ref{fig:circuit_VQA}(b2) has a simple form, $U_b=H^{\otimes{m}}$. Concerning the variational wave function ansatz $|\psi(\vec{\theta})\rangle$, we adapt the form introduced in \cite{liu_variational_2021} to obtain the following ansatz:
\begin{equation}\label{wave_ansatz}
\ket{\psi(\vec{\theta})}=U_{M}(\vec{\beta_{p}})U_{D}(\vec{\gamma_{p}})\cdots U_{M}(\vec{\beta_{1}}) U_{D}(\vec{\gamma_1})\ket{+}^{\otimes m},
\end{equation}
with
\begin{equation}
\begin{aligned}
U_D(\vec{\gamma_{l}})&:=exp(-i\sum_{j=0}^{j=m-2}\gamma_{l}^{j}Z_jZ_{j+1}-i\gamma_{l}^{\,m-1}Z_{m-1}Z_0-i\gamma_l^{\,y}Y_0Y_1),\\
U_M(\vec{\beta_{l}})&:=exp(-i\sum_{j=0}^{j=m-1}\beta_{l}^{j}X_j),
\end{aligned}
\end{equation}
where $\vec{\gamma_l}=(\gamma_l^{\,0},\cdots,\gamma_l^{\,m-1},\gamma_l^{\,y})$, $\vec{\beta_l}=(\beta_l^0,\cdots,\beta_l^{m-1})$, $\vec{\theta}=(\vec{\beta_{1}},\vec{\gamma_{1}},\cdots, \vec{\beta_{p}}, \vec{\gamma_{p}})$ and $p$ is the depth of the variational circuit. Note that this variational wave function ansatz is equivalent to that argued in \cite{liu_variational_2021} when choosing $\gamma_l^{\,0}=\cdots=\gamma_l^{\,m-1}=\gamma_l^{\,y}$ and $\beta_l^0=\cdots=\beta_l^{m-1}$. Therefore, the quantum circuit for realizing the variational ansatz $|\psi(\vec{\theta})\rangle$ can be similarly obtained.

\begin{figure}
\begin{centering}
\includegraphics[scale=0.6]{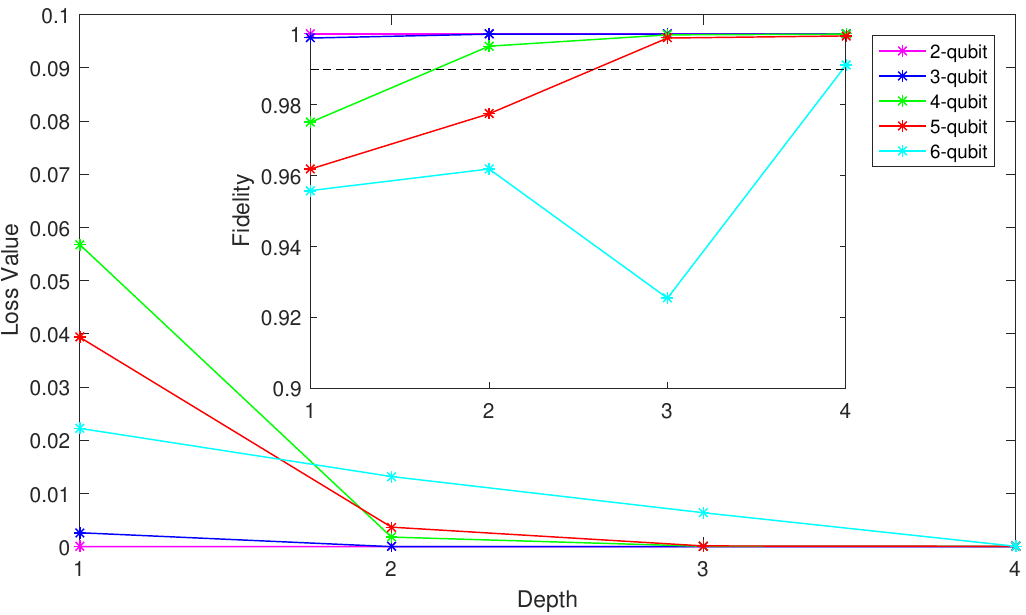}\\
\caption {The value of loss function and the fidelity $|\langle x|\ket{\psi(\vec{\theta})}|^2$ obtained with different depth by using the variational ansatz argued here. The value of the loss function $E(\vec{\theta})$ decreases with the increase of depths, while the fidelity $|\langle x|\ket{\psi(\vec{\theta})}|^2$ generally raises with the increase of depths except for the one case with $m=6$.}
\label{fig:result_ansatz}
\end{centering}
\end{figure}

\begin{figure}
\begin{centering}
\includegraphics[scale=0.6]{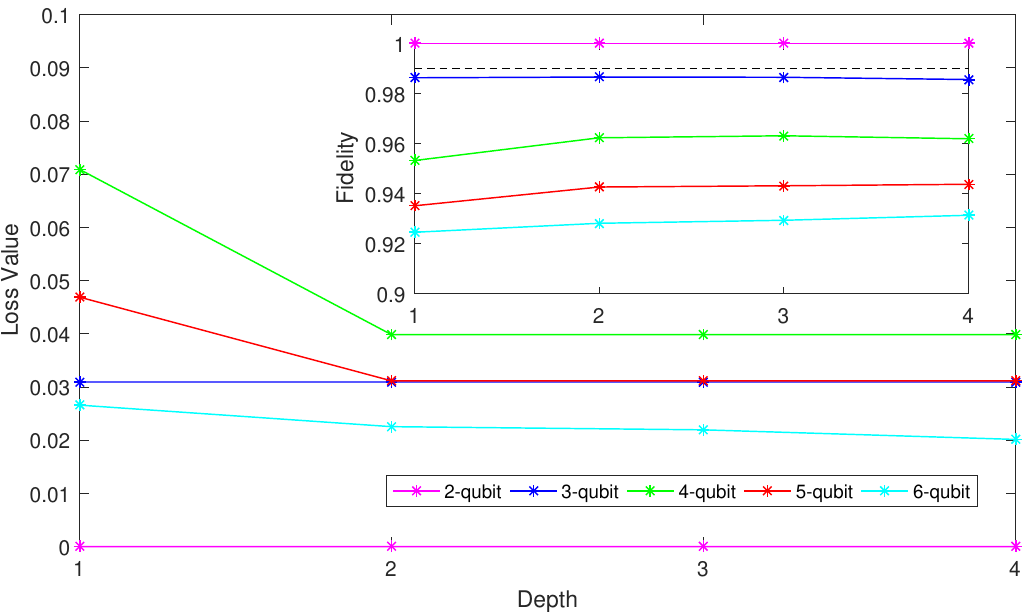}\\
\caption {The result obtained by using the variational ansatz introduced in \cite{liu_variational_2021}. Only the fidelity $|\langle x|\ket{\psi(\vec{\theta})}|^2$ for the size $m=2$ reaches more than 0.99.}
\label{fig:result_ansatz_old}
\end{centering}
\end{figure}

%Here, we focus on the performance of our algorithm on approximating the numerical solution, i.e., the ground state $|x\rangle$ of the Hamiltonian ~\eqref{poisson_hamiltonian}.
In order to avoid convergence to the local minimum, we randomly choose ten angle values from $[0, 2\pi]$ as the initial values for each parameter and execute the VQAs, respectively, to get the optimal parameters by using Gradient-descent methods implemented on classical computers. The advantage of the variational ansatz adopted here compared with the one used in \cite{liu_variational_2021} is obvious (see Fig.~\ref{fig:result_ansatz} and Fig.~\ref{fig:result_ansatz_old}).
%\begin{figure}[!htbp]
%%\centering\includegraphics[width=8cm]{circuit_VQA.pdf}\\
%\centering \includegraphics[width=1\linewidth]{ansatz.pdf}
%\caption{Experimental results of our algorithm. For the sizes argued here, the value of loss function $E(\vec{\theta})$ all decreases with the increase of depths, while the fidelity $|\langle x|\ket{\psi(\vec{\theta})}|^2$ generally raise with the increase of depths except for one case.
%}\label{fig:result_ansatz}
%\end{figure}

From Fig.~\ref{fig:result_ansatz}, we can see that the value of the loss function $E(\vec{\theta})$ decreases with the increase of depths, while the fidelity $|\langle x\ket{\psi(\vec{\theta})}|^2$ generally raises at the same time except for one case. Specifically, for the case $m=6$, we observe that the fidelity decreases when the depth is raised to 3 from 2 in spite of the decreasing of the loss function value at the same time. Even for this worse case, our algorithm still has a good performance, i.e., the fidelity $|\langle x\ket{\psi(\vec{\theta})}|^2$ can reach more than 0.99 with a low-depth VQA for all the sizes considered here.

To illustrate the above worse case, let $H=\sum_{i=0}^{i=2^m-1}\lambda_i |h_i\rangle \langle h_i|$ be a diagonal representation of the argued Hamiltonian $H$. We have $\lambda_0=0<\lambda_1\le\lambda_2\le\cdots\le\lambda_{2^m-1}$ and $|x\rangle=|h_0\rangle$, as the ground state is unique. Assume $\ket{\psi(\vec{\theta})}=\sum_{i=0}^{i=2^m-1}\psi_i|h_i\rangle$. We have
\begin{equation}\label{loss_fedelity}
\begin{aligned}
&E(\vec{\theta})=\sum_{i=1}^{i=2^m-1}\lambda_i|\psi_i|^2,\quad |\langle x\ket{\psi(\vec{\theta})}|^2 = |\psi_0|^2.
\end{aligned}
\end{equation}
From Eq.~\eqref{loss_fedelity}, the decreasing of the loss function $E(\vec{\theta})$ must lead to the increasing of the fidelity $|\langle x|\ket{\psi(\vec{\theta})}|^2$ when $\lambda_1=\lambda_2=\cdots=\lambda_{2^m-1}$, since $E(\vec{\theta})=\lambda_1(\sum_{i=1}^{i=2^m-1}|\psi_i|^2)=\lambda_1(1-|\psi_0|^2)$ in this case. Nevertheless, the fidelity may decrease even if the value of $E(\vec{\theta})$ decreases when the condition $\lambda_1=\lambda_2=\cdots=\lambda_{2^m-1}$ is not satisfied. Suppose $\lambda_j>\lambda_1$ and $\ket{\psi(\vec{\theta_1})}=\sum_{i=0}^{i=2^m-1}\tilde{\psi}_i|h_i\rangle$ with $\tilde{\psi}_i=\psi_i$ for $i\neq 1,j$. We have
\begin{equation}
\begin{aligned}
&E(\vec{\theta_1})-E(\vec{\theta})=\lambda_1(|\tilde{\psi}_1|^2-|\psi_1|^2)+\lambda_j(|\tilde{\psi}_j|^2-|\psi_j|^2),\\
&|\langle x\ket{\psi(\vec{\theta_1})}|^2-|\langle x\ket{\psi(\vec{\theta})}|^2=(|\psi_1|^2-\tilde{\psi}_1|^2)+(|\psi_j|^2-\tilde{\psi}_j|^2).
\end{aligned}
\end{equation}
Set $E(\vec{\theta_1})-E(\vec{\theta})<0$ and $|\langle x\ket{\psi(\vec{\theta_1})}|^2-|\langle x\ket{\psi(\vec{\theta})}|^2<0$. We have
$\lambda_1(|\tilde{\psi}_1|^2-|\psi_1|^2)<\lambda_j(|\psi_j|^2-|\tilde{\psi}_j|^2)$ and $|\tilde{\psi}_1|^2-|\psi_1|^2>|\psi_j|^2-|\tilde{\psi}_j|^2$.
Thus, the worse case occurs when
\begin{equation}
\begin{aligned}
&|\psi_j|^2-|\tilde{\psi}_j|^2>0,~~~
1<\frac{\tilde{\psi}_1|^2-|\psi_1|^2}{|\psi_j|^2-|\tilde{\psi}_j|^2}
<\frac{\lambda_j}{\lambda_1},
\end{aligned}
\end{equation}
which may be satisfied when $\lambda_j>\lambda_1$. Intuitively, the larger the value $\frac{\lambda_j}{\lambda_1}$, the more likely the worse case occurs. Moreover, we classically compute the eigenvalues of the target Hamilonian with $m=2,3,4,5,6$, and obtain that the ratio of the maximum eigenvalue to the subminimum eigenvalue can reach $160000$ for $m=6$. Therefore, it is understood that the worse case occurs at $m=6$ in our numerical experiments. We emphasize that the form of loss function and the eigenvalue distribution of the target Hamiltonian $H$ both contribute to the occurrence of the worse case argued above. How to avoid or reduce the occurrence of such worse case is also an interesting problem.

%Therefore, we may avoid this phenomenon by in two ways. The one way is choosing a more suitable form of the loss function or optimizing process to reducing the occurrence of the phenomenon. The other way is to reconstruct the target Hamiltonian for which the ratio of the largest eigenvalue to the lesser eigenvalue is as close to 1 as possible under the requirement that the solution to target linear system can be encoded to the its unique ground state. How to
%Choosing $\tilde{\psi}_1>\psi_1$ and $\tilde{\psi}_2<\psi_1$
%the minimum depth of the variational circuit that is required to guarantee the 0.99 fidelity is $1,1,2,3,4$, respectively. Interestingly, for the case $m=2$, i.e., the number of qubits is 2, we observed that the fidelity decreases when depth raises to 3 from 2 while the value of loss function decreases at the same time.

% $|x\rangle$ by estimating the fidelity of the quantum solution to the ground state for the Hamiltonian \eqref{poisson_hamiltonian}, namely, $|\langle x|\psi_1(\vec{\theta})\rangle|^2$ or $|\langle x|\psi_2(\vec{\theta})\rangle|^2$.

%which is proportional to the exact solution of the linear system generated by the discretization of the Poisson equation, by estimating

\section{Conclusion}
\label{conclusion}
In this paper, we have focused on solving the one-dimensional Poisson equations with different boundary conditions and the $d$-dimensional Poisson equations with Dirichlet boundary conditions by using VQAs.
Given the Hamiltonian whose ground state encodes the solution to the discretized Poisson equation with coefficient matrix $A$, we have effectively evaluated the loss function by utilizing the sparsity of $A$. In detail, we have written the loss function in terms of the operator $\sigma_x\otimes A$ with $\sigma_x$ denoting the standard Pauli operator, as $\sigma_x\otimes A$ can be directly decomposed into a sum of Hermitian, one-sparse, and self-inverse operators. Then, for the one-dimensional Poisson equations with different boundary conditions and for the $d$-dimensional Poisson equations with Dirichlet boundary conditions, we have decomposed $\sigma_x\otimes A$ into a sum of at most 7 and $(4d+1)$ Hermitian, one-sparse and self-inverse operators, respectively. Finally, the quantum circuits have been explicitly constructed to evaluate efficiently the loss function.

We emphasize that our algorithm to evaluate the loss function is effective as the number of the decomposition terms is only polynomial of $d$ and is independent of $n$. Therefore, our algorithm greatly reduces the number of measurements compared with the algorithm presented in \cite{liu_variational_2021}. This advantage is particularly outstanding in dealing with the linear systems with large dimension $n$. In addition, the number of the ancilla qubits required for realizing the quantum circuits corresponding to the decomposition terms is at most $\log_2n+2$ and is independent of $d$. The circuit complexity is also polynomial of $\log_2n$ and independent of $d$, which implies that the benefit of our algorithm is still obvious for the high-dimensional Poisson equations.

It is noteworthy that the decomposition method and the quantum circuit design presented here are still suitable and efficient for the linear systems with Hermitian and sparse coefficient matrices satisfying $a_{i,i+c}=a_{c}$ for all $c=0,1,\cdots,n-1$ and $i=0,\cdots,n-1-c$, where $a_{i,i+c}$ denotes the element of the coefficient matrices. Additionally, inspired by our method, for the coefficient matrices which slightly violate the above conditions one may also find the corresponding decomposition.

Numerically, we have adapted the variational ansatz introduced in the previous works to test our algorithm for the one-dimensional Poisson equations with Dirichlet boundary conditions. It has been shown that the value of the loss function decreases with the increase of depths, while the fidelity $|\langle x\ket{\psi(\vec{\theta})}|^2$ generally raises at the same time except for one case. Specifically, for the case $m=6$, the fidelity decreases when the depth is raised to 3 from 2 in spite of the decreasing of the loss function at the same time. This poor phenomenon is attributed to the form of the loss function and the eigenvalue distribution of the target Hamiltonian $H$. How to avoid or reduce the occurrence of such poor phenomenon is also an interesting problem. We emphasize that, even for such worse case, our algorithm still has such a good performance that the fidelity $|\langle x\ket{\psi(\vec{\theta})}|^2$ can reach more than 0.99 with low-depth VQAs for all sizes.

\begin{acknowledgements} This work is supported by National Natural Science Foundation of China Grants No.12075159 and No.12171044, Beijing Natural Science Foundation Grant No.Z190005, and the Academician Innovation Platform of Hainan Province.
\end{acknowledgements}

\begin{appendices}
\appendix
\label{appendix}
\section{Quantum circuits for the Hermitian, one-sparse, and self-inverse terms used in Sec.~\ref{sec:VQA_one_dimention} and Sec.~\ref{sec:VQA_d_dimention}}
\begin{figure}
\begin{centering}
\includegraphics[scale=0.4]{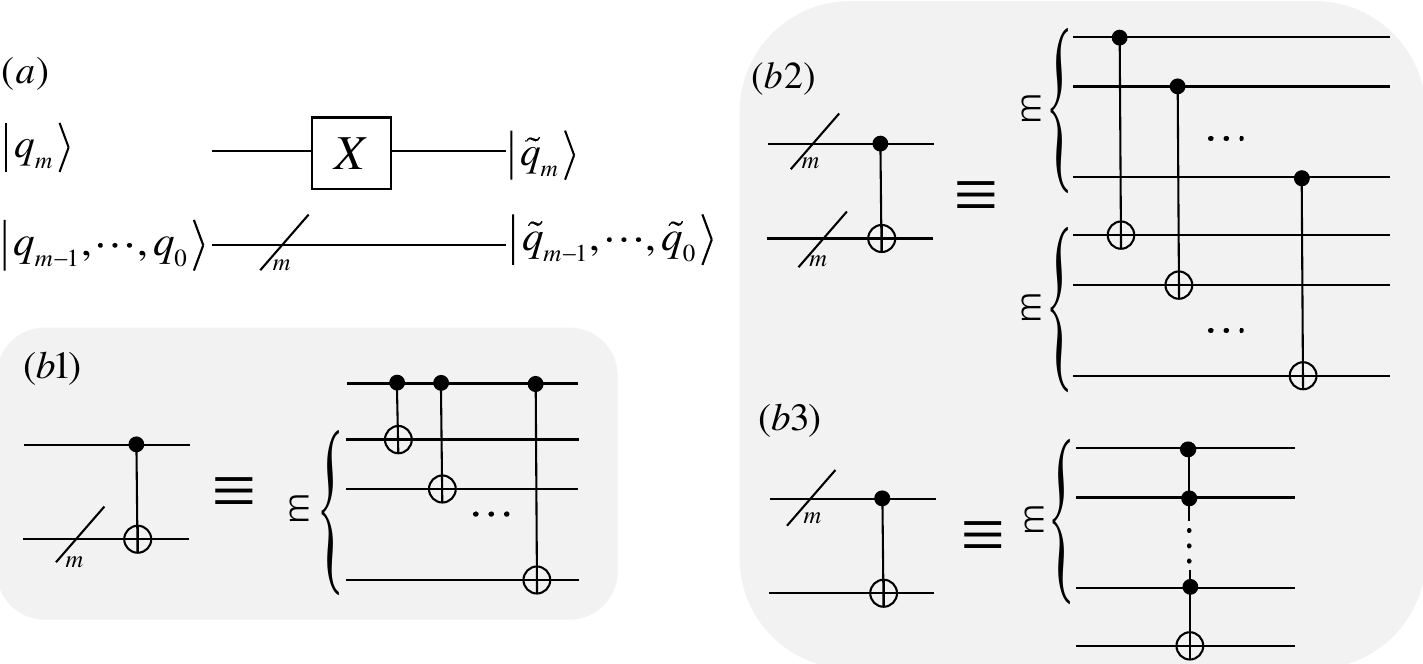}\\
\caption {The quantum circuit for the decomposition term $G_0$ and some simplified circuit denotations.
($a$) The quantum circuit for the decomposition term $G_0$.
(b1)-(b3) Some simplified circuit denotations used in this paper.}
\label{fig:decomposition_term_circuitG0}
\end{centering}
\end{figure}

\renewcommand{\thesubsection}{A}
\label{sec:appendix_A}
The realization of the quantum circuits for the decomposition operators is the key in our algorithm. It is observed from \eqref{one_dimension_decomposition_effects} and \eqref{d_dimension_decomposition_effects} that the quantum circuits corresponding to the operators $G_0$ and $G_{d,0}$ can be designed without any ancilla qubits [see Fig.~\ref{fig:decomposition_term_circuitG0}($a$) and Fig.~\ref{fig:decomposition_term_circuitG_{d,0}}].
\begin{figure}
\begin{centering}
\includegraphics[scale=0.4]{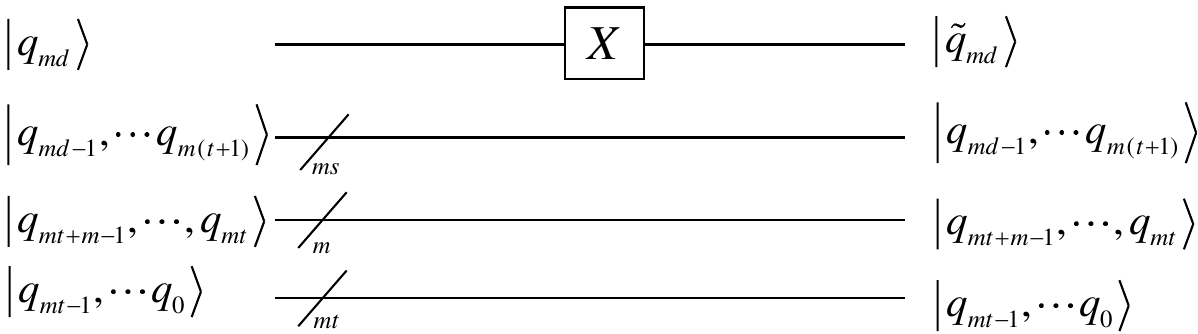}\\
\caption {The quantum circuit for the decomposition term $G_{d,0}$.}
\label{fig:decomposition_term_circuitG_{d,0}}
\end{centering}
\end{figure}

Let $|0\rangle_{\text{phase}}$ and $|0_m\cdots 0_10_0\rangle_{a}$
be two ancilla registers used to control the global phase and to realize the query to the Hermitian, one-sparse and self-inverse operator $G$. Namely, for a given computational basis state $|x\rangle$, $G_{xy_x}|y_x\rangle_{a}$ corresponds to the state obtained after executing $G$ on $|x\rangle$, where $y_x$ is the column index of the nonzero entry in row $x$ of $G$ and $G_{xy_x}=\pm 1$ is the corresponding element. The register $|0\rangle_{\text{phase}}$ is used to determine whether $G_{xy_x}=-1$ and to determine whether $y_x=x$. The register $|0_m\cdots 0_10_0\rangle_{a}$ is used to store the column index $y_x$.

With respect to the operators $G_0^{f-}$ and $G_0^{l-}$, the column index $y(q_m\cdots q_0)$ of the nonzero entry in the row $q_m\cdots q_0$ can be obtained by flipping the state of $|q_m\rangle$. Moreover, only the entries in rows $0\cdots 0$ and $10\cdots0$ of $G_0^{f-}$ (or in rows $01\cdots 1$ and $11\cdots 1$ of $G_0^{l-}$) are $-1$ instead of $1$. Therefore, only the ancilla register $|0\rangle_{\text{phase}}$ is used to construct the corresponding quantum circuits [see Fig.~\ref{fig:decomposition_term_circuitG_0^{f-}_G_0^{l-}}($a$)]. For the operators $G_1^{f\pm}$, the quantum states $|q_m\rangle$ and $|q_0\rangle$ are both flipped when $q_{m-1}=\cdots= q_1=0$, and the global phase is set to be $-1$ ($+1$) at the same time after executing $G_1^{f-}$ ($G_1^{f+}$) on the computational basis state $|q_{m}\cdots q_0\rangle$. Thus, the ancilla register $|0\rangle_{\text{phase}}$ is used to design the circuits for $G_1^{f\pm}$ (see Fig.~\ref{fig:decomposition_term_circuitG_1^{f}_G_1^{l}}). Similarly, the quantum circuits for the operators $G_1^{l\pm}$ are constructed (see Fig.~\ref{fig:decomposition_term_circuitG_1^{f}_G_1^{l}}).
\begin{figure}
\begin{centering}
\includegraphics[scale=0.4]{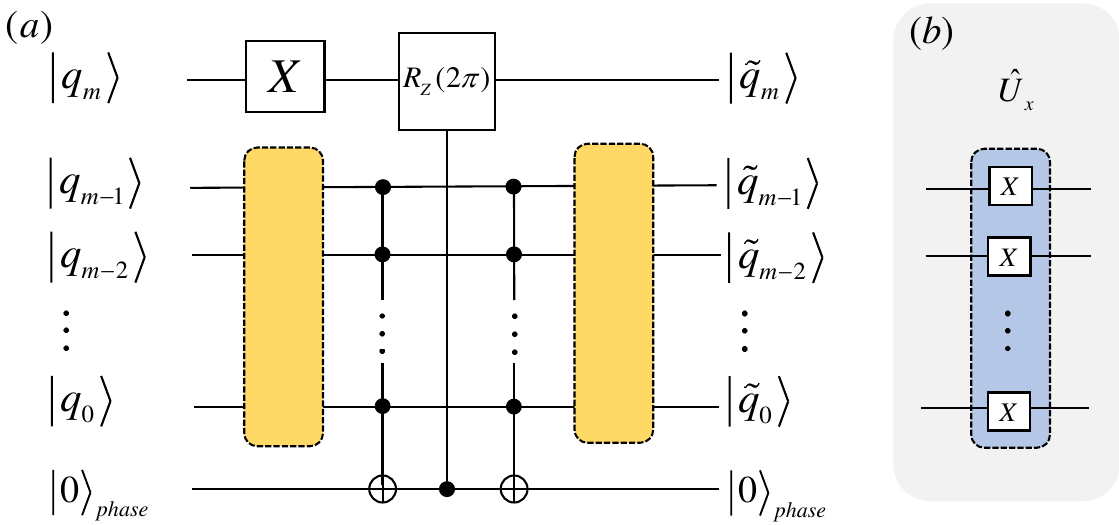}\\
\caption {The quantum circuit for the decomposition terms $G_0^{f-}$ and $G_0^{l-}$.
($a$) The quantum circuit corresponds to the term $G_0^{f-}$ when the yellow boxes are both replaced by the operator $\hat{U_x}$ shown in (b), and corresponds to the term $G_0^{l-}$ when the yellow boxes do not include any operation.
(b) The explicit form of operator $\hat{U}_x$ used in this paper.}
\label{fig:decomposition_term_circuitG_0^{f-}_G_0^{l-}}
\end{centering}
\end{figure}
\begin{figure}
\begin{centering}
\includegraphics[scale=0.4]{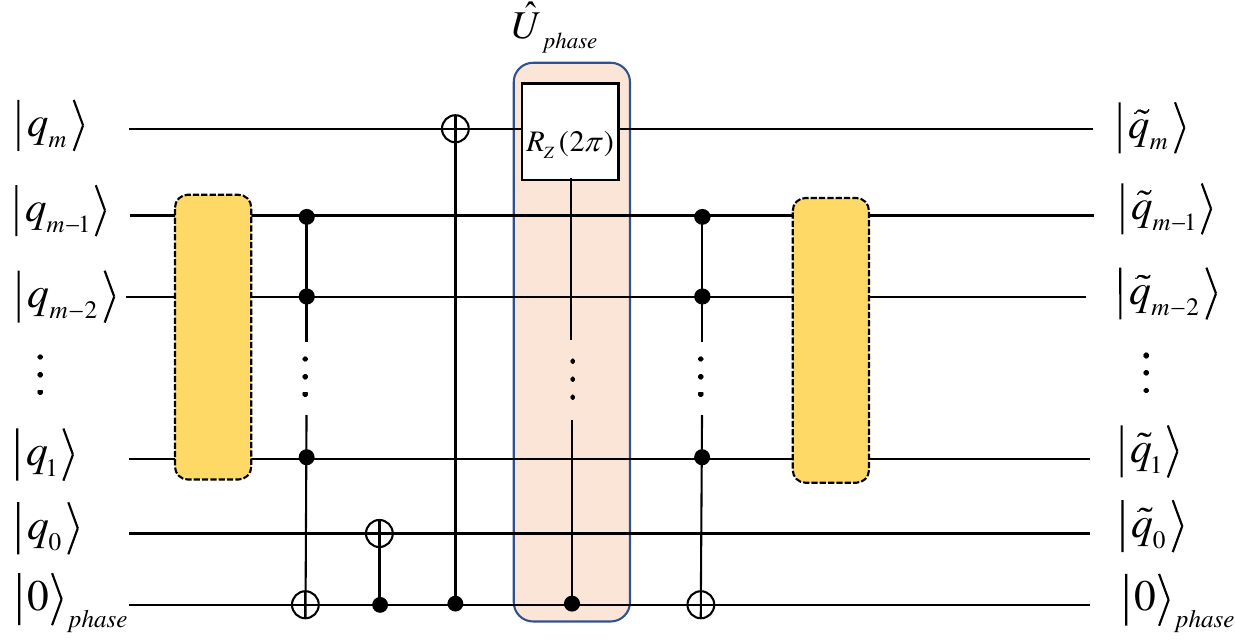}\\
\caption {The quantum circuits for the decomposition terms $G_1^{f\pm}$ and $G_1^{l\pm}$. This circuit corresponds to the terms $G_1^{f\pm}$ when the yellow boxes are both replaced by the operator $\hat{U_x}$ shown in Fig.~\ref{fig:decomposition_term_circuitG_0^{f-}_G_0^{l-}}(b), and corresponds to the terms $G_1^{l\pm}$ when the yellow boxes do not include any operation. Whether the operation $\hat{U}_{phase}$ is removed or not determines whether this circuit corresponds to the term $G_1^{f+}$ ($G_1^{l+}$) or the $G_1^{f-}$ ($G_1^{l-}$).}
\label{fig:decomposition_term_circuitG_1^{f}_G_1^{l}}
\end{centering}
\end{figure}

 Aiming to give the general steps to realize the quantum circuits for other operators used in Sec.~\ref{sec:VQA_one_dimention} and Sec.~\ref{sec:VQA_d_dimention}, we first analyze the operators $G_{d,+2^{mt}}^{\pm}$ and $G_{d,-2^{mt}}^{\pm}$.
It is observed from \eqref{d_dimension_decomposition_effects} that the operators $G_{d,+2^{mt}}^{\pm}$ and $G_{d,-2^{mt}}^{\pm}$ only work on the quantum states $|q_{md}\rangle$ and $|q_{mt+m-1}\cdots q_{mt}\rangle$. Let $|g'\rangle$, with $g'=\sum_{i=0}^{m-1}q_{mt+i}2^i+2^mq_{md}$, denote the computational basis state $|q_{md}q_{mt+m-1}\cdots q_{mt}\rangle$. The actions of the operators $G_{d,+2^{mt}}^{\pm}$ and $G_{d,-2^{mt}}^{\pm}$ can be shown as
\begin{equation}
\begin{aligned}\label{d_dimension_decomposition_effects_rewritten}
&G_{d,+2^{mt}}^{\pm}:|g'\rangle \rightarrow \pm \prod_{i=0}^{i=m-1}q_{mt+i}|g'\rangle+(1-\prod_{i=0}^{i=m-1}
q_{mt+i})|g'+2^{m}+1\rangle,~g'=0,\cdots,2^{m}-1,\\
&\qquad\quad\ \ \  |g'\rangle \rightarrow \pm \prod_{i=0}^{i=m-1}(1-q_{mt+i})|g'\rangle
+\left(1-\prod_{i=0}^{i=m-1}(1-q_{mt+i})\right)|g'-2^{m}-1\rangle,~g'=2^{m},\cdots,2^{m+1}-1,\\
&G_{d,-2^{mt}}^{\pm}:|g'\rangle \rightarrow \pm \prod_{i=0}^{i=m-1}(1-q_{mt+i})|g'\rangle
+\left(1-\prod_{i=0}^{i=m-1}(1-q_{mt+i})\right)|g'+2^{m}-1\rangle,~g=0,\cdots,2^{m}-1\\
&\qquad\quad\ \ \  |g'\rangle \rightarrow \pm \prod_{i=0}^{i=m-1}q_{mt+i}|g'\rangle
+\left(1-\prod_{i=0}^{i=m-1}q_{mt+i}\right)|g'-2^{m}+1\rangle,~ g=2^{m},\cdots,2^{m+1}-1,
\end{aligned}
\end{equation}
which are of the similar forms to the ones given by the operators $G_{+1}^{\pm}$ and $G_{-1}^{\pm}$ [see  \eqref{one_dimension_decomposition_effects}]. Similar results can be obtained for the operators $G_{+2}^{\pm}$ and $G_{-2}^{\pm}$. Thus, we focus on illustrating the design for the quantum circuits corresponding to the operators $G_{+1}^{\pm}$ and $G_{-1}^{\pm}$ below.

Similar to the steps introduced in \cite{kirby_variational_2021}, we apply the following steps to design the quantum circuits for the operators $G_{+1}^{\pm}$ and $G_{-1}^{\pm}$:
\begin{equation}\label{circuit_steps}
\begin{aligned}
|q_m\cdots q_0\rangle_q |0_m\cdots 0_0\rangle_{a}|0\rangle_{\text{phase}}&\xrightarrow{\hat{O}_G}|q_m\cdots q_0\rangle_q |y(q_m\cdots q_0)\rangle_{a}|G_{q,y(q)}\rangle_{\text{phase}} \\
&\xrightarrow{\hat{U}_{phase}}G_{q,y(q)}|q_m\cdots q_0\rangle_q |y(q_m\cdots q_0)\rangle_{a}|G_{q,y(q)}\rangle_{\text{phase}}\\
&\xrightarrow{\text{swap }q,a}G_{q,y(q)}|y(q_m\cdots q_0)\rangle_{q}|q_m\cdots q_0\rangle_a |G_{q,y(q)}\rangle_{\text{phase}}\\
&\xrightarrow{(\hat{O}_G)^{-1}}G_{q,y(q)}|y(q_m\cdots q_0)\rangle_{q}|0_m\cdots 0_0\rangle_a |0\rangle_{\text{phase}},
\end{aligned}
\end{equation}
where $|q_m\cdots q_0\rangle_q $ denotes any input computational basis state, $y(q_m\cdots q_0)$ is the column index of the nonzero entry in row $q_m\cdots q_0$ of $G_{+1}^{\pm}$ or $G_{-1}^{\pm}$, and the value of the single nonzero entry is denoted by $G_{q,y(q)}$. It is straightforward to see that the phase-control operator $\hat{U}_{phase}$ can be removed when the nonzero entries are all $1$, namely, the circuit for $G_{+1}^{+}$ ($G_{-1}^{+}$) is similar to that for $G_{+1}^{-}$ ($G_{-1}^{-}$) except that the phase-control operator $\hat{U}_{phase}$ is removed. Therefore, we focus on analyzing the circuit design for the operators $G_{+1}^{-}$ and $G_{-1}^{-}$ below.

From \eqref{circuit_steps}, we see that the circuit design for the operator $\hat{O}_G$ is the key. In order to simplify the quantum circuit of $\hat{O}_G$, we take into account the symmetry of $G_{+1}^{-}$ and $G_{-1}^{-}$, namely, the process $|1q_{m-1}\cdots q_0\rangle\xrightarrow{G}|y(1q_{m-1}\cdots q_0)\rangle$ can be realized through the following steps:
\begin{equation}
\begin{aligned}\label{1_to_0}
|1q_{m-1}\cdots q_0\rangle&\xrightarrow{\text{flip all qubits}}|0(1-q_{m-1})\cdots (1-q_0)\rangle\xrightarrow{G}|y\left(0(1-q_{m-1})\cdots (1-q_0)\right)\rangle\xrightarrow{\text{flip all qubits}}|y(1q_{m-1}\cdots q_0)\rangle.
\end{aligned}
\end{equation}
$|0q_{m-1}\cdots q_0\rangle\xrightarrow{G}|y(1q_{m-1}\cdots q_0)\rangle$ can also be realized in a similar way,
\begin{equation}\label{0_to_1}
\begin{aligned}
|0q_{m-1}\cdots q_0\rangle&\xrightarrow{\text{flip all qubits}}|1(1-q_{m-1})\cdots (1-q_0)\rangle\xrightarrow{G}|y\left(1(1-q_{m-1})\cdots (1-q_0)\right)\rangle\xrightarrow{\text{flip all qubits}}|y(0q_{m-1}\cdots q_0)\rangle,
\end{aligned}
\end{equation}
where $G$ represents either $G_{+1}^{-}$ or $G_{-1}^{-}$.

With respect to the operator $G_{+1}^{-}$, its action on the state $|1q_{m-1}\cdots q_0\rangle$ can be transformed to that on the state $|0(1-q_{m-1})\cdots (1-q_0)\rangle$ by using the process \eqref{1_to_0}. It can be seen from \eqref{one_dimension_decomposition_effects} that the state $|0q_{m-1}\cdots q_0\rangle$ is changed to be $|1(q_{m-1}\cdots q_0)+1\rangle$ after executing the operator $G_{+1}^{-}$ when $q_{m-1},\cdots,q_0$ are not all 1, and $|0q_{m-1}\cdots q_0\rangle$ is changed to be $-|0q_{m-1}\cdots q_0\rangle$ when $q_{m-1}=\cdots=q_0=1$, where '+' denotes the binary addition. Thus, we can design the operator $\hat{U}_{+1}$ to realize the binary addition [see Fig.~\ref{fig:decomposition_term_circuitG_{+1}} (b)]. The register $|0\rangle_{\text{phase}}$ is flipped to $|1\rangle_{\text{phase}}$ when $q_{m-1}=\cdots=q_0=1$ so that the effect on the state $|0q_{m-1}\cdots q_0\rangle$, where $q_{m-1},\cdots,q_0$ are not all 1, can be realized with the help of the operator $\hat{U}_{+1}$ controlled by $|0\rangle_{\text{phase}}$, while the effect on the state $|01\cdots 1\rangle$ can be realized with the help of the operator $R_{z}(2\pi)$ controlled by $|1\rangle_{\text{phase}}$. Consequently, the quantum circuit for $G_{+1}^{-}$ is designed in Fig.~\ref{fig:decomposition_term_circuitG_{+1}}.

A similar argument applies to the circuit design for the operators $G_{-1}^{-}$ except that the action of $G_{-1}^{-}$ on the state $|0q_{m-1}\cdots q_0\rangle$ is transformed to that on the state $|1(1-q_{m-1})\cdots (1-q_0)\rangle$ by using the process \eqref{0_to_1}. The corresponding quantum circuits can be found in  Fig.~\ref{fig:decomposition_term_circuitG_{-1}}.

Combining the discussion \eqref{d_dimension_decomposition_effects_rewritten} and the circuits for $G_{\pm1}^{\pm}$, all the quantum circuits $G_{+2}^{\pm}$, $G_{-2}^{\pm}$, $G_{d,+2^{mt}}^{\pm}$ and $G_{d,-2^{mt}}^{\pm}$ are also constructed (see Fig.~\ref{fig:decomposition_term_circuitG_{+2}}, Fig.~\ref{fig:decomposition_term_circuitG_{-2}}, Fig.~\ref{fig:decomposition_term_circuitG_{+2mt}} and Fig.~\ref{fig:decomposition_term_circuitG_{-2mt}}, respectively).
\begin{figure}
\begin{centering}
\includegraphics[scale=0.4]{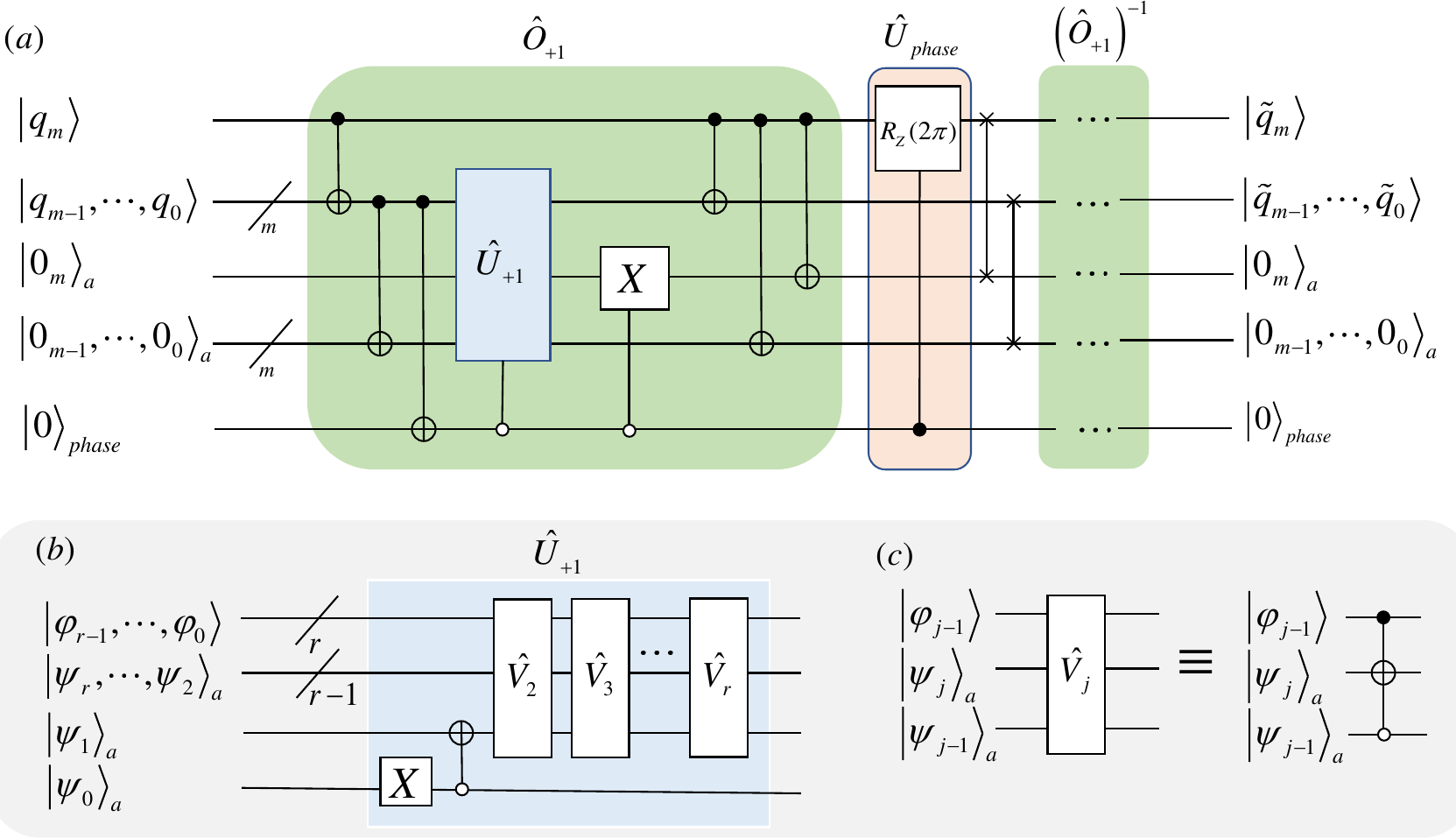}\\
\caption {The quantum circuit for the decomposition terms $G_{+1}^{\pm}$.
($a$) The circuit corresponds to the term $G_{+1}^{+}$ when the operator $\hat{U}_{phase}$ is removed, and corresponds to the term $G_{+1}^{-}$ when $\hat{U}_{phase}$ is retained.
(b) The quantum circuit for the operator $\hat{U}_{+1}$ used in this paper.
(c) The explicit form of the operator $\hat{V}_j$ included in the quantum circuit of $\hat{U}_{+1}$.}
\label{fig:decomposition_term_circuitG_{+1}}
\end{centering}
\end{figure}
\begin{figure}
\begin{centering}
\includegraphics[scale=0.4]{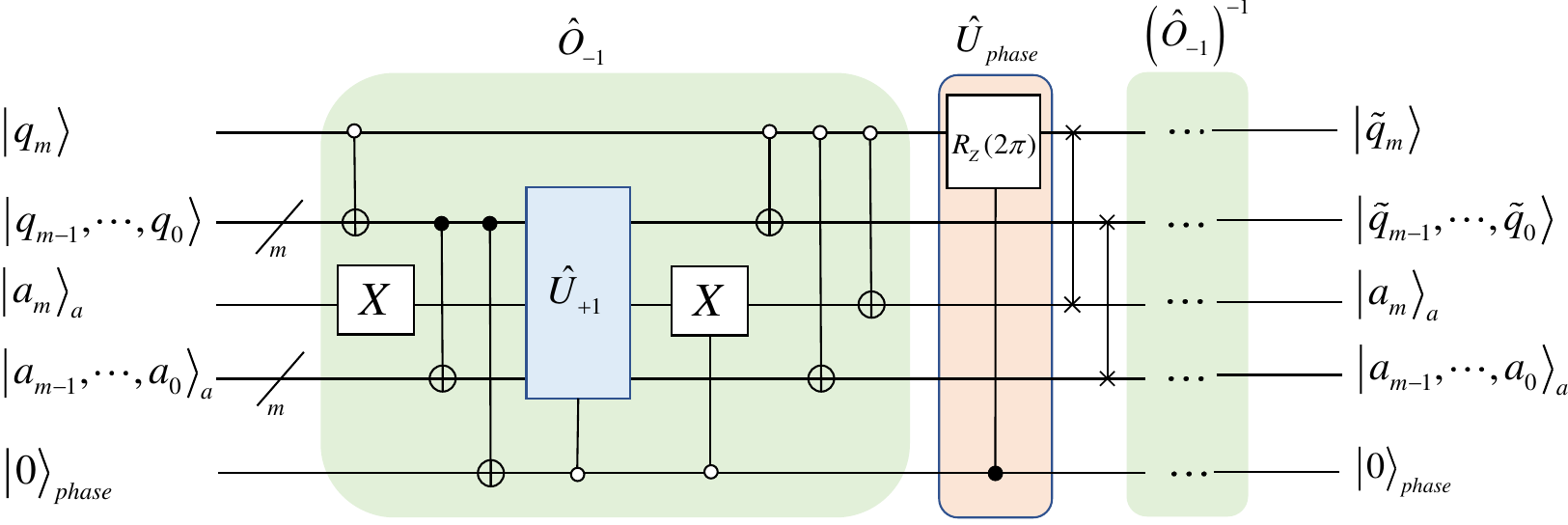}\\
\caption {The quantum circuits for the decomposition terms $G_{-1}^{+}$ and $G_{-1}^{-}$ when the operator $\hat{U}_{phase}$ is removed and retained, respectively.}
\label{fig:decomposition_term_circuitG_{-1}}
\end{centering}
\end{figure}
\begin{figure}
\begin{centering}
\includegraphics[scale=0.4]{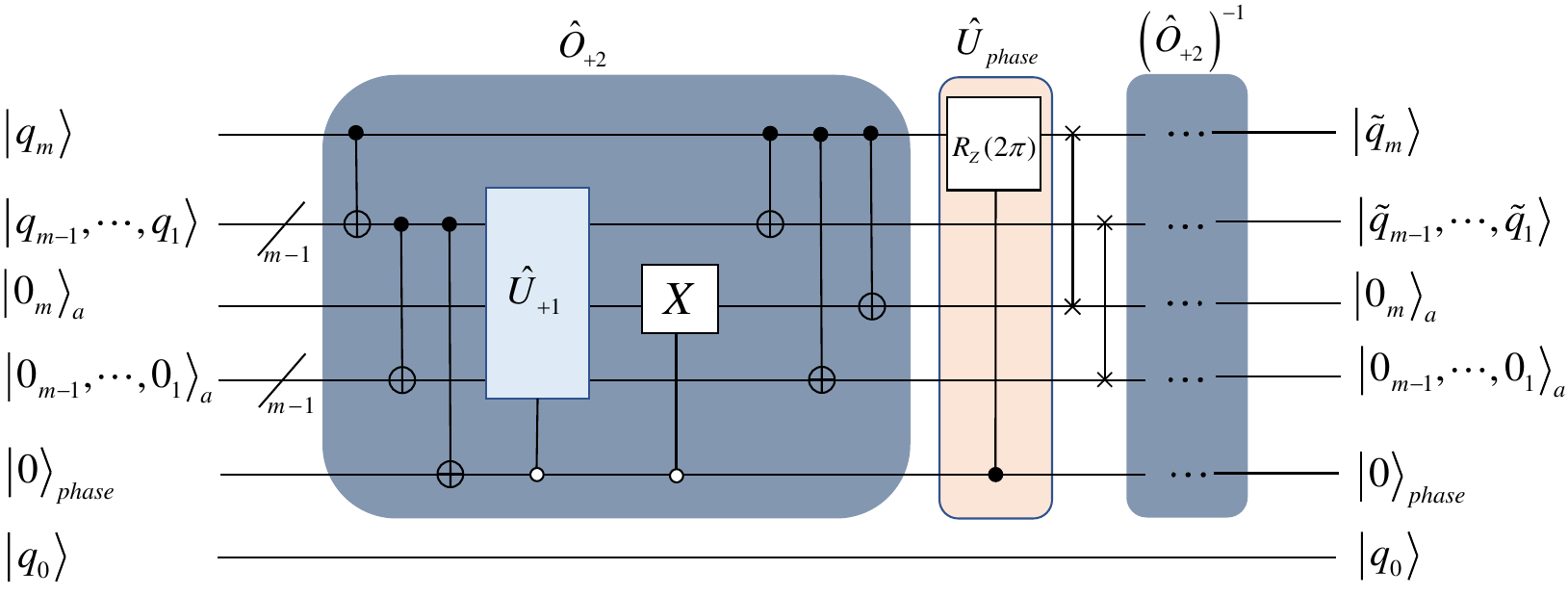}\\
\caption {The quantum circuits for the decomposition terms $G_{+2}^{+}$ and $G_{+2}^{-}$ when the operator $\hat{U}_{phase}$ is removed and retained, respectively.}
\label{fig:decomposition_term_circuitG_{+2}}
\end{centering}
\end{figure}
\begin{figure}
\begin{centering}
\includegraphics[scale=0.4]{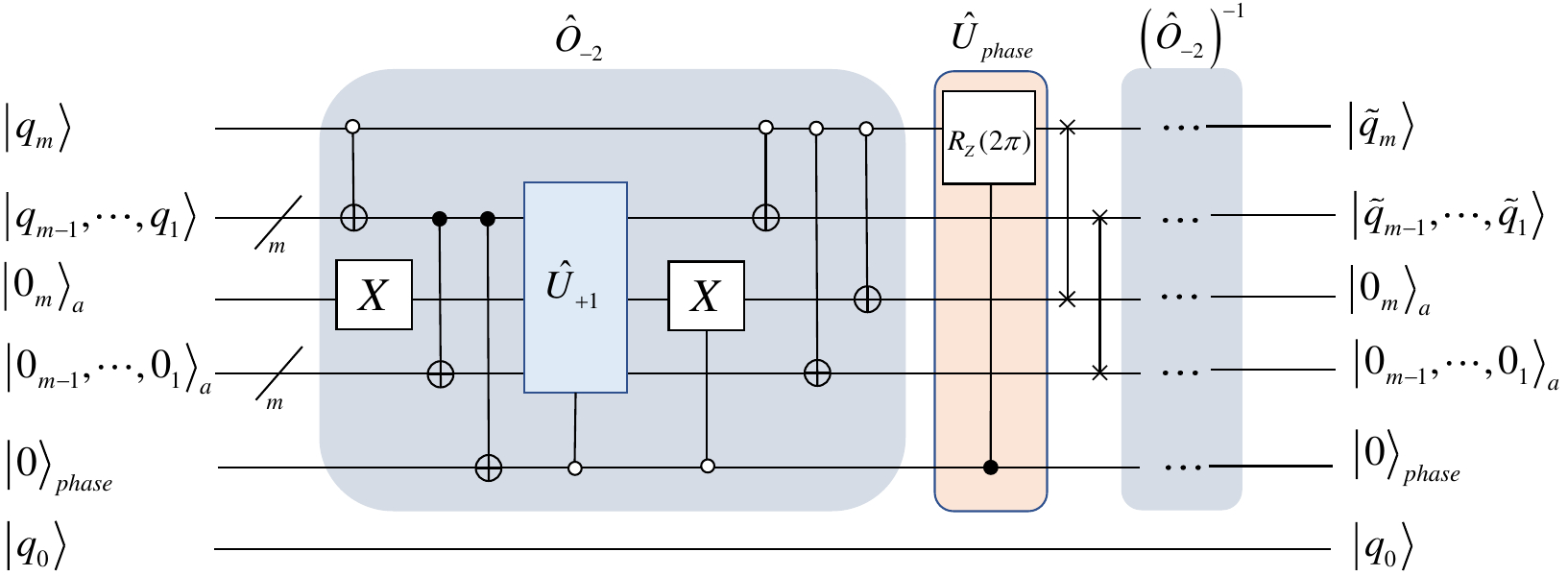}\\
\caption {The quantum circuits for the decomposition terms $G_{-2}^{+}$ and $G_{-2}^{-}$ when the operator $\hat{U}_{phase}$ is removed and retained, respectively.}
\label{fig:decomposition_term_circuitG_{-2}}
\end{centering}
\end{figure}
\begin{figure}
\begin{centering}
\includegraphics[scale=0.4]{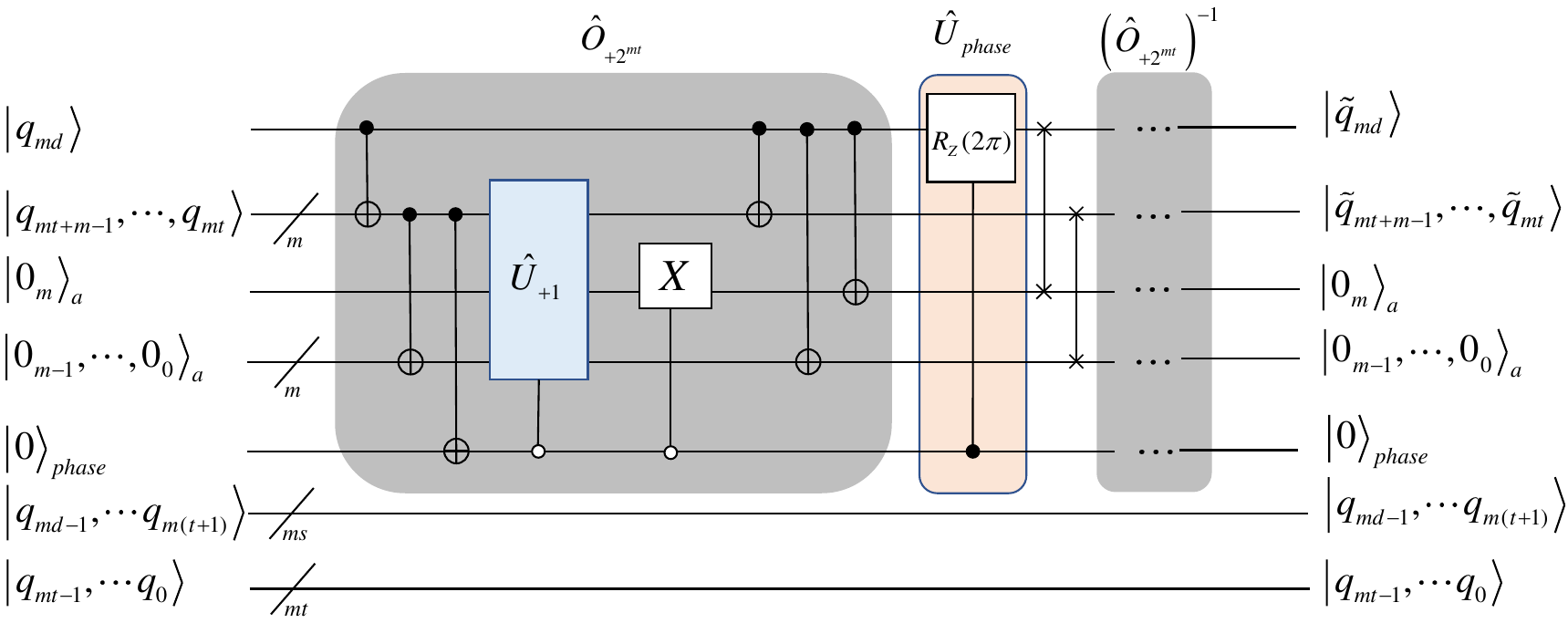}\\
\caption {The quantum circuits for the decomposition terms $G_{d,+2^{mt}}^{+}$ and $G_{d,+2^{mt}}^{-}$ when the operator $\hat{U}_{phase}$ is removed and retained, respectively.}
\label{fig:decomposition_term_circuitG_{+2mt}}
\end{centering}
\end{figure}
\begin{figure}
\begin{centering}
\includegraphics[scale=0.4]{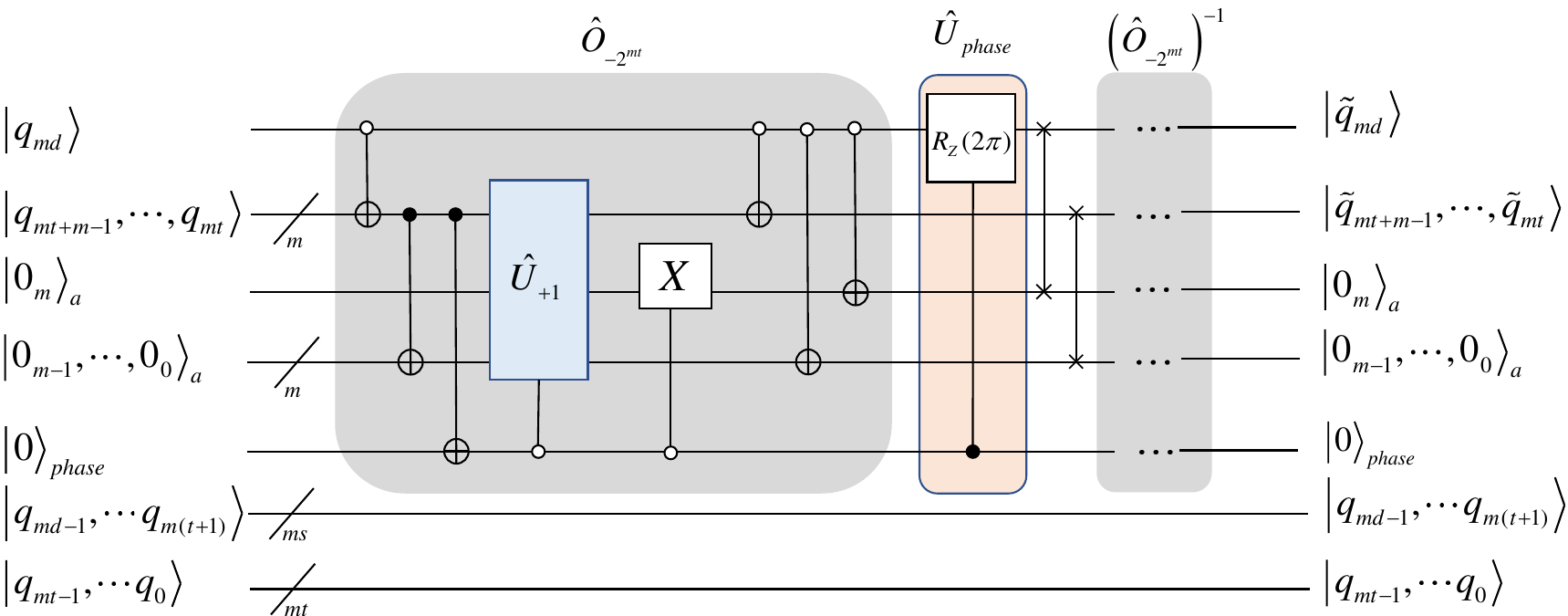}\\
\caption {The quantum circuits for the decomposition terms $G_{d,-2^{mt}}^{+}$ and $G_{d,-2^{mt}}^{-}$ when the operator $\hat{U}_{phase}$ is removed and retained, respectively.}
\label{fig:decomposition_term_circuitG_{-2mt}}
\end{centering}
\end{figure}
\end{appendices}

\end{document}